\documentclass[11pt]{article}
\usepackage{fullpage}
\usepackage{subfig}
\usepackage[usenames,dvipsnames]{xcolor}
\usepackage[linktocpage=true,
	pagebackref=true,colorlinks,
	linkcolor=BrickRed,citecolor=blue]
	{hyperref}
	\usepackage{latexsym,graphicx,epsfig,color}
\usepackage{amsfonts,amssymb,amsmath,amsthm,amstext}
\usepackage{enumitem}
\setitemize{itemsep=2pt,topsep=0pt,parsep=0pt}
\usepackage{url,setspace}
\usepackage{multirow}
\usepackage{rotating}
\usepackage{makeidx}
\usepackage{tikz}
\usepackage{xspace}
\usepackage{algorithm,algpseudocode}
\usepackage{bm}
\usepackage{changepage} 	
\usepackage{thmtools,thm-restate}
\usepackage{cleveref}

\newcommand{\IGNORE}[1]{}
\allowdisplaybreaks

\usetikzlibrary{decorations.markings}
\usetikzlibrary{arrows}
\tikzstyle{block}=[draw opacity=0.7,line width=1.4cm]
\tikzstyle{graphnode}=[circle, draw, fill=black!20, inner sep=0pt, minimum width=6pt]
\tikzstyle{point}=[circle, draw, fill=black!30, inner sep=0pt, minimum width=1pt]
\tikzstyle{input}=[rectangle, draw, fill=black!75,inner sep=3pt, inner ysep=3pt, minimum width=4pt]
\tikzstyle{unmatched}=[graphnode,fill=black!0]
\tikzstyle{shaded}=[graphnode,fill=black!20]
\tikzstyle{matched}=[graphnode,fill=black!100]  	
\tikzstyle{matching} = [ultra thick]
\tikzset{
    >=stealth',
    pil/.style={
           ->,
           thick,
           shorten <=2pt,
           shorten >=2pt,}
}
\tikzset{->-/.style={decoration={
  markings,
  mark=at position .5 with {\arrow{>}}},postaction={decorate}}}

\makeindex

\newtheorem{theorem}{Theorem}[section]
\newtheorem{claim}[theorem]{Claim}
\newtheorem{proposition}[theorem]{Proposition}
\newtheorem{lemma}[theorem]{Lemma}

\newtheorem{definition}[theorem]{Definition}

\newtheorem{remark}[theorem]{Remark}

\usepackage{thmtools,thm-restate} 

\newcommand{\E}{\mathbb{E}}

\newcommand{\OPT}{\textsc{OPT}}

\newcommand{\calX}{\mathcal{X}}

\def\eps {\epsilon}

\def \reals {\mathbb{R}}

\newcommand{\e}{\epsilon}
\newcommand{\mtop}[1]{\textrm{Top-}#1}
\newcommand{\R}[0]{\ensuremath \mathbb{R}}
\newcommand{\W}{\mathcal{W}}

\newcommand{\ones}{\bm{1}}
\newcommand{\zero}{\bm{0}}
\newcommand{\ip}[2]{\langle #1, #2\rangle}

\newcommand{\one}{\mathbf{1}\xspace}
\newcommand{\poly}{\operatorname{poly}\xspace}

\newif\ifsubmit
\submitfalse

\newcounter{note}[section]

\ifsubmit
    \newcommand{\focs}[1]{{#1}}
    
    \newcommand{\snote}[1]{}
    \newcommand{\tnote}[1]{}
    \newcommand{\mnote}[1]{}
\else
    \newcommand{\focs}[1]{{#1}}
    
    \newcommand{\snote}[1]{\refstepcounter{note}$\ll${\bf Sahil~\thenote:}
      {\sf \color{blue}  #1}$\gg$\marginpar{\tiny\bf SS~\thenote}}
    \newcommand{\tnote}[1]{\refstepcounter{note}$\ll${\bf Thomas~\thenote:}
      {\sf \color{blue}  #1}$\gg$\marginpar{\tiny\bf TK~\thenote}}
    \newcommand{\mnote}[1]{\refstepcounter{note}$\ll${\bf Marco~\thenote:}
      {\sf \color{blue}  #1}$\gg$\marginpar{\tiny\bf MM~\thenote}}
\fi

\newcommand{\Load}{\ensuremath{\Lambda}\xspace}
\newcommand{\load}{\ensuremath{\Lambda}\xspace}
\newcommand{\OPTBwK}{\ensuremath{\textsc{OPT}_{\textsc{BwK}}}\xspace}
\newcommand{\Regret}{\textsc{Regret}}

\newcommand{\loadBal}{\textsc{GenLoadBal}\xspace}
\newcommand{\vecSched}{\textsc{VecSched}\xspace}
\newcommand{\onGAP}{\textsc{OnGAP}\xspace}
\newcommand{\BwK}{\textsc{BwK}\xspace}
\newcommand{\BwVC}{\textsc{BwVC}\xspace}

\newcommand{\alg}{\textrm{Alg}\xspace}

\newcommand{\ballopt}{\textsf{Ball-Optimization}\xspace}
\newcommand{\norm}{\textsf{norm}}

\newcommand{\vertm}[1]{{\left\vert\kern-0.25ex\left\vert\kern-0.25ex\left\vert #1     \right\vert\kern-0.25ex\right\vert\kern-0.25ex\right\vert}}
\newcommand{\argmin}{\textrm{argmin}}

\title{Online and Bandit Algorithms Beyond $\ell_p$ Norms
}

	\author{Thomas Kesselheim\thanks{
	(thomas.kesselheim@uni-bonn.de)
 Institute of Computer Science,	    University of Bonn. 
    }
	\and Marco Molinaro\thanks{
         (mmolinaro@microsoft.com)
         Microsoft Research and PUC-Rio. 
    }
	\and Sahil Singla\thanks{
        (ssingla@gatech.edu)
        School of Computer Science,
        Georgia Tech.
        }
}

\def\hasmain{}

\date{ \today}

\begin{document}
\maketitle

\begin{abstract}
{\medskip 

Vector norms play a fundamental role in computer science and optimization, so there is an ongoing effort to generalize existing algorithms to settings beyond  $\ell_\infty$ and $\ell_p$ norms. 
We show that  many online and bandit applications  for general norms  admit good algorithms as long as the norm can be approximated by a function that is ``gradient-stable'', a notion that we introduce. Roughly it says that the gradient of the function should not drastically decrease (multiplicatively) in any component as we increase the input vector. 
We prove that several families of norms, including all monotone symmetric norms, admit a gradient-stable  approximation, giving us the first  online and bandit algorithms for these norm families.

\smallskip
In particular, our notion of gradient-stability gives $O\big(\log^2 (\text{dimension})\big)$-competitive algorithms for the symmetric norm generalizations of   Online Generalized Load Balancing and Bandits with Knapsacks. Our techniques extend to applications beyond  symmetric norms
as well, e.g., to   Online Vector Scheduling and to Online Generalized Assignment with Convex Costs. Some key properties underlying our applications that are implied by gradient-stable approximations  are  a ``smooth game inequality'' and an approximate converse to Jensen's inequality.


}
\end{abstract}

\thispagestyle{empty}
\clearpage

\setcounter{tocdepth}{2}
{\small
   \tableofcontents
} 

\thispagestyle{empty}

\clearpage

\setcounter{page}{1}


\ifx \hasmain \undefined
  \input{preamble}
  \begin{document}
\fi

\section{Introduction}
	Many fundamental problems in optimization and computer science involve norms of vectors, implicitly or explicitly. 
	A classic example is Load Balancing or machine scheduling, where we need to assign jobs to machines in order to minimize the  maximum load on a machine: This objective is nothing but the $\ell_\infty$-norm of the vector of loads incurred by each machine. Other examples that explicitly deal with norms include Nearest Neighbors Problem where the distance between points is a norm and Discrepancy Minimization where we minimize a norm of the signed sum of vectors. There are many other applications that do not explicitly involve a norm but can be phrased as such. For instance,  observe that every set of non-negative vectors $\mathcal{C}$ defines a norm: $\|x\|_{\mathcal{C}} := \sup_{c \in \mathcal{C}} \sum_i c_i |x_i|$. An important example of this observation is that the class of XOS  functions\footnote{A functions $f : 2^V \rightarrow \R$ over a finite ground set $V$ is XOS iff it can be written as $f(S) := \max_{c \in \mathcal{C}} \sum_{e \in S} c(e)$ for some set  $\mathcal{C}$ of non-negative weight vectors.}, which contains submodular functions and has  applications in mechanism design and submodular optimization~\cite{FeigeV06,FeldmanGL15,CaiZ17,AssadiKS21}, can be viewed as norms.	
	
	For some more concrete examples of norms and its applications, we have the classic $\ell_p$-norms (including $\ell_\infty$) $\|x\|_p := (\sum_i x_i^p)^{1/p}$. Another example is the $\mtop{k}$-norm, which is the sum of the largest $k$ absolute values of the coordinates of the vector; these provide another important family of interpolations between the $\ell_1$ ($k=d$) and $\ell_\infty$ ($k=1$) norms. More broadly we have the \emph{ordered norms}, namely those of the form $\|x\| = \sum_i w_i |x|^\downarrow_i$ for non-negative weights $w_1 \geq w_2 \geq \ldots \geq w_d \in \R_+$,	where $|x|^\downarrow$ is the vector with the absolute value of the coordinates of $x$ sorted in non-increasing order; see the books~\cite{locationTheory1,locationTheory2} for their applications in location theory, and \cite{kMedian1,kMedian2} for their applications in $k$-clustering. Another rich subset of norms are \emph{Orlicz norms}, which are defined as
	$\|x\|_f := \inf\big\{ \lambda > 0 ~\big|~ \sum_i f\big(\frac{x_i}{\lambda}\big) \le 1 \big\}$ for any convex function $f$ with $f(0) =0$
	. 
	E.g., the $\ell_p$-norm is the special case when $f(z) = z^p$. See~\cite{orliczRegression1,orliczRegression2} for recent applications of these norms to regression. 
	
	We note that all the  norms mentioned in the previous paragraph are \emph{monotone}, namely $\|u\| \ge \|v\|$ whenever $u \ge v \ge 0$, and  \emph{symmetric}, namely they are invariant to the permutation of the coordinates of the space. Indeed, most ``naturally occurring'' norms satisfy these properties. See~\cite{CS-STOC19} for further examples  and \cite[Chapter IV]{BhatiaBook} for basic  properties of monotone symmetric norms. 

Since norms play an important role in many areas, there have been several recent works trying to understand more general norms in different settings. To mention a few examples, there are works on offline Load Balancing and $k$-clustering with symmetric monotone norms~\cite{CS-STOC19,offLoadBalNorms2,loadBalEasy}, stochastic Load Balancing and spanning trees (also with symmetric monotone norms)~\cite{stochLoadBalNorms,loadBalEasy}, nearest-neighbor search~\cite{ANNNorms,ANNHolder} (first for symmetric monotone norms, and then for general norms), 
linear regression with Orlicz norms~\cite{orliczRegression1,orliczRegression2}, and mean estimation with statistical queries~\cite{meanEstimationNorms}. However, all these works are either in an offline or a stochastic setting. Despite of all this progress, we are not aware of general techniques to obtain similar results for \textbf{online} (adversarial) problems. For example, the results from~\cite{CS-STOC19,offLoadBalNorms2,loadBalEasy} on offline Load Balancing and related problems (the closest to the problems we consider) are obtained by rounding the solution of fairly complex convex programs, and it is unclear how to solve and round them online. This motivates our main question regarding online algorithms and online learning: 

\vspace{6pt}
\centerline{ \emph{How to design algorithms for online problems with general monotone norms?}}


\subsection{Summary of Our Results}

	
	In this paper we introduce the notion of \textbf{gradient-stable approximations} of norms (\Cref{def:gradStable})  and use it to obtain new results in a unified manner for several classic online problems involving norms, including \textbf{Online Generalized  Load Balancing} and \textbf{Bandits with Knapsacks}. The idea is to approximate the norm by a function that has the property that its gradient at $x \in \R_+^d$ does not to decrease drastically in any component when increasing $x$ by a small amount. We show that if an approximation with this gradient-stable property has a multiplicative error $\alpha$ and an additive error $\gamma$ then we can obtain online algorithms with competitive ratio $O(\alpha + \gamma)$.
	
	Crucially, one of our key technical contributions is to show that \textbf{every} monotone symmetric norm admits a gradient-stable approximation with a multiplicative error $\alpha = O(\log d)$ and an additive error $\gamma = O(\log^2 d)$. This  gives us $O(\log^2 d)$-approximations with monotone symmetric norms, the first such general results for the problems considered. We also prove that gradient-stability works well under norm compositions, which allow us to extend the results beyond symmetric norms. 
	



\subsection{Results for Online Algorithms}






\paragraph{Online Generalized Load Balancing.} In the fundamental problem of scheduling on unrelated machines to minimize makespan (see the books~\cite{borodin2005online,BuchbinderNaor-Book09}), the goal is to minimize the $\ell_\infty$-norm of the vector of all machine loads. The more general case where the norm is the $\ell_p$-norm $\|x\|_p := (\sum_i x_i^p)^{1/p}$ has also been studied since the 70's \cite{chandra,cody}, since in many applications they better capture how well-balanced an allocation is~\cite{awerbuch}. These are some of the special cases of the \emph{Online Generalized Load Balancing} problem~\cite{simultaneousLoadBal}.



 In Online Generalized Load Balancing there are $m$ machines, and $T$ jobs come one-by-one. Each job can be processed by the machines in $k$ different ways, so the $t$-th job has a $m \times k$ non-negative matrix $C^{(t)}$ whose column $j$ gives the loads $(C^{(t)}_{1j}, C^{(t)}_{2j}, \ldots, C^{(t)}_{mj})$ that the machines incur if the job is processed with option $j$. When the $t$-th job arrives, the algorithm needs to select a processing option for it (namely a vector $x^{(t)} \in \{0,1\}^k$ with exactly one 1) based only on the jobs seen thus far. To measure the quality of the solution, it is given a norm $\|\cdot\|$ over $\R^m$. The goal is to minimize the norm of the total load incurred on the machines, namely $\| \sum_{t = 1}^T C^{(t)} x^{(t)} \|$. The performance of the algorithm is compared against the offline optimal solution $\OPT := \min_{x_*^{(1)}, \ldots, x_*^{(T)}} \|\sum_{t = 1}^T C^{(t)} x^{(t)}_*\|$, and it is said to be a  $\beta$-approximation if its total load is at most $\beta\OPT$.

For the classic setting of unrelated scheduling with makespan minimization ($\ell_\infty$-norm, diagonal matrices $C^{(t)}$), traditional algorithms give an $O(\log m)$-approximation~\cite{AzarNR95,AspnesAFPW-JACM97}, and this was generalized for the setting of $\ell_p$-norms by~\cite{awerbuch,Caragiannis-SODA08} to obtain an $O\big(\min\{p, \log m\}\big)$-approximation for all $p \in [1,\infty]$ (all these results are optimal, up to constant factors). For the problem of online routing ($\ell_\infty$-norm, but the columns of $C^{(t)}$ form paths on a graph)~\cite{AspnesAFPW-JACM97}, there is also an optimal $O(\log m)$-approximation. The results in \cite{simultaneousLoadBal} give a $O\big(\min\{p, \log m\}\big)$-approximation for the problem with $\ell_p$-norm (and arbitrary matrices $C^{(t)}$, thus generalizing the above). Optimal results are also known for specifically structured norms, in which case the problem is known as \emph{Online Vector Scheduling} (more on this below). 
For more general norms, only the \textbf{offline} version of the problem was recently solved:~\cite{CS-STOC19,offLoadBalNorms2,loadBalEasy} obtained constant-factor approximations for every monotone symmetric norm. 

Our main result for this problem is the following (see \Cref{def:gradStable} for gradient-stability):
	
	\begin{theorem}[Load Balancing] \label{thm:loadBalInformal}
		For  Online Generalized Load Balancing, if the norm $\|\cdot\|$ admits a $\frac{1}{4}$-gradient-stable approximation with error $(\alpha,\gamma)$, where $\alpha$ is the multiplicative error and $\gamma$ is the additive error, then a greedy algorithm obtains an $O(\alpha + \gamma)$ competitive ratio. \focs{Moreover, this greedy algorithm is efficient given value and gradient oracle access to the $\frac14$-gradient stable approximation.}
	\end{theorem}
	
    In the appendix (\Cref{lemma:smoothLp}), we show that a known approximation of $\ell_p$-norms for $p \in [1,\infty]$ is indeed a $\frac{1}{4}$-gradient-stable approximation with error $(1,\min\{p-1, \log m\})$. So, \Cref{thm:loadBalInformal} directly recovers the optimal $O\big(\min\{p, \log m\}\big)$-approximations from~\cite{awerbuch,Caragiannis-SODA08,simultaneousLoadBal}. Using instead our new approximation of symmetric norms (stated in \Cref{thm:symm}) allows us to obtain the following polylog approximation for \textbf{any} monotone symmetric norm.
	
	
 
	
\begin{theorem} \label{thm:loadBalSymm}
		Consider the Online Generalized Load Balancing problem with any monotone symmetric norm $\|\cdot\|$. Then, there is an algorithm that obtain an $O(\log^2 m)$ competitive ratio. \focs{Moreover, this algorithm is efficient given \ballopt oracle\footnote{We use the definition in \cite{CS-STOC19} whereby \ballopt oracle allows us to compute  $\max_{v: \|v\|\leq 1} \langle x, v \rangle$ for any vector $x \in \R^d$ with a single oracle call.} access to the norm $\|\cdot\|$.} 
	\end{theorem}

We note that previously no polylog competitive algorithms were known  even for $\mtop{k}$ norm or ordered  norms.
Since even for the $\ell_\infty$-norm there is an $\Omega(\log m)$ lower bound on the approximation factor~\cite{AzarNR95},  \Cref{thm:loadBalSymm} is optimal up to the exact polylog.


	\paragraph{Applications Beyond Symmetric Norms.} Our insights (especially the general \Cref{thm:loadBalInformal}) have implications beyond symmetric norms. As one example, we consider the \textbf{Online Vector Scheduling} problem. In brief, this is a problem similar to Online Load Balancing but where  each machine has $r$ resources (e.g., a server with CPU, network, and disk); when a job is assigned to a machine, it loads these resources in different ways. For each resource $i$, there is an ``inner'' norm that measures the distribution of the load on this resource across the machines. The algorithm's goal is to minimize the largest (makespan) of these inner norms. (See \Cref{sec:vecSched} for a formal definition of the problem.)  


	Following up on previous works, \cite{vecSched} gave an optimal $O(\log m + \log r)$-approximation algorithm when the inner norms are $\ell_p$'s.  We are able to go beyond their results and obtain a polylog approximation for Online Vector Scheduling with any monotone symmetric inner norms. Although  it is not difficult to cast Online Vector Scheduling as a special case of the Online Generalized Load Balancing with a ``nested norm'', this  norm is \textbf{not symmetric} even if the inner norms are. Nonetheless, we show that in this case we can still obtain low-error gradient-stable approximations, which yields the following result.  
	
    
	\begin{theorem} \label{thm:vecSchedSym}
		Consider the Online Vector Scheduling problem where the inner norms are arbitrary monotone symmetric norms. Then, there is an  $O(\log^2 m + \log r)$-competitive algorithm. \focs{Moreover, this algorithm is efficient given \ballopt oracle access to the inner norms.}
	\end{theorem}
	
	We remark that in the special case where the inner norms are $\ell_p$, our result improves to an $O(\log m + \log r)$-approximation, which recovers the result from~\cite{vecSched}.
	
	\medskip

    \focs{Another related application is the \textbf{Online $k$-Sided Placement problem}~\cite{ksided}; for simplicity we briefly describe the case $k=2$. There are $k=2$ sets of machines offering different services (e.g. storage and processing). Each machine has a capacity. Jobs come online and need to be assigned to one storage and one processing machine. When assigned to a machine pair, a job adds to the load of these machines and also incurs a cost, both load and cost depending on the job and machine pair. In its minimization version, the goal is to find an assignment minimizing the total cost while respecting the machine capacities. In the offline case,~\cite{ksided} gives an algorithm that obtains optimal cost while violating the capacities by a factor of at most $(k+1)$ (such violation is required for any non-trivial approximation in the cost). While the paper also consider an online maximization version of the problem, nothing is known for this minimization version. 
    
    
    As in Online Vector Scheduling (plus using a guessing of $\OPT$) one can see that this problem is a special case of Online Load Balancing with a non-symmetric nested norm. Then using the same ideas, we can obtain an online algorithm that has cost at most $O(\log m)$ times  $\OPT$ and violates the capacities by at most a factor of $O(\log m)$, where $m$ is again the total number of machines.

    
    }

    \medskip

	As another application of our techniques, we consider a problem where the objective  is not even a norm. In \textbf{Online Generalized Assignment with Convex Costs} (\onGAP), we again have  $m$ machines and $T$ online jobs with $k$ processing options each. However, now each job has both a \emph{cost} and a \emph{duration} that it will incur over the machines. There are two monotone convex functions $f_{cost}, f_{dur}: \R^m_+ \rightarrow \R_+$ that measure the total cost and total duration that a schedule induces on the machines. The goal is to find a schedule of jobs that  minimizes the sum $f_{cost} + f_{dur}$. This is an online and a generalized version of the classic Generalized Assignment Problem of Shmoys-Tardos~\cite{ST}, which is a building block of algorithms on a myriad of applications (for a sample see~\cite{usesGAP,PTASMultiKnap,anupamSLB,stochLoadBalM} and the book~\cite{handbookApprox}). 
 
	In~\cite{raviConvexPD} the authors consider a special case of \onGAP where $f_{cost}$ is the $\ell_p$-norm raised to the power of $p$ and $f_{dur}$ is the $\ell_1$-norm. They were motivated by applications to energy efficient scheduling and routing. In this special case they obtain an optimal $O(p^p)$-approximation.
 	Here we can generalize their result to arbitrary symmetric $f_{cost}$'s and $f_{dur}$'s with a competitive ratio that depends on ``growth order'', a standard dependence when working with non-homogeneous functions~\cite{BGMS11,huangKim,cvxPDFOCS,sodaMixed}. Formally, a function $f$ has \emph{growth order at most $p$} if $f(\alpha x) \le \alpha^p f(x)$ for all $\alpha \ge 1$ and  $x \ge 0$. E.g., the function $\|\cdot\|_p^p$ (used in  previous work), and more generally polynomials of degree $p$ with non-negative coefficients, have growth order at most~$p$. 
	
	\begin{restatable}{theorem}{onlineGAP} \label{thm:GAP}
	  Consider the problem \onGAP with functions $f_{cost}, f_{dur}$ that are convex, monotone, and symmetric. Then there is an $O(\log^{2p} m)$-competitive 
	  algorithm for this problem, where $p$ is the maximum of the growth order of $f_{cost}$ and $f_{dur}$. \focs{Moreover, this algorithm is efficient given  \ballopt access to the  norms $\| \cdot \|_{f_{cost}}$ and $\|\cdot \|_{f_{dur}}$, where for a convex function $f$ its norm is $\| x\|_{f}:= \inf\big\{ \lambda > 0 \mid f \big(\frac{x}{\lambda}\big) \le 1\big\}$.} 
	\end{restatable}
	
	The lower bound from Online Load Balancing in $\ell_\infty$-norm~\cite{AzarNR95} implies an $\Omega(\log^p m)$ lower bound for this problem, so again our result is tight up to a factor of two in the exponent.

	

\subsection{Results for Bandit Algorithms}

Next we discuss our results for online learning problems with bandit (partial) feedback.

\paragraph{Bandits with Knapsacks.} In this problem we are given a budget $B \geq 0$, a norm $\| \cdot \|$, and a set  of $n$ actions. In each  time step $t \in [T]$, we take one of these actions  $x^{(t)} \in \{e_1, \ldots, e_n \}$, and then receive a \emph{scalar} reward $\langle r^{(t)}, x^{(t)} \rangle$ and incur a $d$-dim \emph{vector} cost $ C^{(t)} x^{(t)} $,  where $r^{(t)} \in [0,1]^{1 \times n}$ is a row vector of rewards and $C^{(t)} \in [0,1]^{d \times n}$ is a cost matrix. Moreover, we also receive a  bandit feedback  $ C^{(t)} x^{(t)}$ and $\langle r^{(t)}, x^{(t)} \rangle$, i.e., we do not get to see the entire cost matrix or reward vector but only the part pertaining to the played action after having made the choice.  We assume there is a \emph{null action} in the action set that gives $0$ reward and $\zero$ vector cost, which allows us to skip some time steps. The goal is to maximize the total  reward received while  ensuring that the $\| \cdot \|$ norm of the total cost vector is  less than $B$. After exhausting the budget, we are only allowed to choose the null action, and thus we obtain no further reward.
The benchmark for \BwK is any fixed fractional selection  $x^\ast \in \Delta_n$ of  actions, where $\Delta_n$ is the $n-1$ dimensional simplex $\{x \in [0,1]^n \mid \sum_i x_i \leq 1\}$.

This problem was first introduced in \cite{BKS-JACM18} in the special case where the rewards and cost vectors are drawn i.i.d.  and the norm is $\ell_\infty$. 
Since then the problem has been generalized to adversarial rewards/costs~\cite{ISSS-FOCS19} and the norm has been generalized to  $\ell_p$-norms~\cite{KS-COLT20}, obtaining tight $O(\log d)$-approximations when the optimal value is known. These works were motivated by applications such dynamic item pricing, repeated auctions, and dynamic procurement where the bandit actions consume a budget besides giving a reward.
Our result vastly  generalizes the  previous settings to arbitrary norms that admit gradient-stable approximations. 

\begin{restatable}{theorem}{thmbwK}
\label{theorem:bwk}
Consider the  Bandits with Knapsacks problem for adversarial arrivals with a norm $\|\cdot\|$. Suppose $\|\cdot\|$ admits a $\frac14$-gradient-stable approximation with error $(\alpha,\gamma)$. Furthermore, let $B \geq 4 \cdot (\alpha + \gamma) \cdot \| \ones \|$. Then, there exists an algorithm that takes $\OPTBwK$ as its input and obtains reward at least
\[
\Omega\Big(\frac{1}{\alpha + \gamma}\OPTBwK\Big) - O\Big(\frac{\OPTBwK \cdot \| \one \| }{(\alpha + \gamma) \cdot B}\Big)\cdot \textsc{Regret}.
\]
with probability $1-p$, where $\textsc{Regret} = O(\sqrt{T n \log(n/p)})$ and $p \in [0,1]$ is a parameter. \focs{Moreover, this algorithm is efficient given gradient oracle access to  $\frac14$-gradient stable approximation of $\| \cdot \|$.}
\end{restatable}

In particular, for monotone symmetric norms $\|\cdot\|$ this gives the guarantee $$\Omega\Big(\frac{1}{\log^2 d}\OPTBwK\Big) - O\Big(\frac{\OPTBwK \cdot \| \one \| }{\log^2 d \cdot B}\Big)\cdot \textsc{Regret},$$ which generalizes the previous results while still obtaining a $\poly\!\log d$ approximation. The assumption that $\OPTBwK$ is known can be removed  at a further multiplicative loss of $\Theta(\log T)$, which is known to be unavoidable~\cite{ISSS-FOCS19}.




Our techniques  to prove this theorem using gradient-stable norm approximations also apply to other bandit problems, e.g, the following ``Bandits with Vector Costs'' problem.


\paragraph{Bandits with Vector Costs.}
This problem is a natural generalization of the classic adversarial bandits problem~\cite{DBLP:journals/siamcomp/AuerCFS02} where we incur vector costs instead of scalar costs. The goal is to minimize a given norm of the total cost vector. Formally, we are given a  set of $n$ actions and in each  time step $t$ we take one of these actions  $x^{(t)} \in \{e_1, \ldots, e_n \}$. After taking the action we  incur a \emph{vector} cost $ C^{(t)} x^{(t)} $, where $C^{(t)} \in [0,1]^{d \times n}$ is a cost matrix, and receive a bandit feedback  $ C^{(t)} x^{(t)} $, i.e., only for the played action. The goal of the algorithm is to minimize a given norm $\|\cdot \|$ of its total cost vector compared to that of a fixed  fractional selection $x^\ast \in \Delta_n$ over the actions.

In the special case of $\ell_p$-norms, this problem was introduced in \cite{KS-COLT20}. They obtained a tight $O(\min\{p,\log d\})$ approximation in this setting. Our work generalizes this result to all norms admitting gradient-stable approximations. 


\newcounter{banditsVC}
\setcounter{banditsVC}{\value{theorem}}

\begin{theorem} 
	Consider the problem Bandits with Vector Costs with a norm $\|\cdot\|$. If $\|\cdot\|$ admits a $\frac{1}{4}$-gradient-stable approximations with error $(\alpha,\gamma)$, then there exists an algorithm that guarantees $\lVert \sum_{s=1}^T C^{(t)} \cdot x^{(t)} \rVert = O((\alpha + \gamma) \cdot \lVert \sum_{s=1}^T C^{(t)} \cdot x^\ast \rVert + \alpha \cdot \lVert \ones \rVert \cdot \textsc{Regret})$ with probability $1-p$, where $\textsc{Regret} = O(\sqrt{T n \log(n/p)})$ and $p \in [0,1]$ is a parameter. \focs{Moreover, this algorithm is efficient given value and gradient oracle access to this $\frac14$-gradient stable approximation to $\| \cdot \|$.}
 \end{theorem}

 Again, in the case of monotone symmetric norms $\|\cdot\|$ this gives the guarantee $O(\log^2 d \cdot \lVert \OPT \rVert + \log d \cdot \lVert \ones \rVert \cdot \textsc{Regret})$, generalizing the result of~\cite{KS-COLT20} while still obtaining a $\poly\!\log d$ approximation.





\subsection{Gradient-Stable Approximations and their Key Properties} \label{sec:introSGI}
 
 	As mentioned above, the central concept underlying our algorithms is that of approximating norms with functions whose gradients are stable. Namely, we introduce the following definition.

\begin{definition}[Gradient-Stable Approximation] \label{def:gradStable}
	We say that a norm $\|\cdot\|$ admits a $\delta$-gradient-stable approximation with error $(\alpha, \gamma$)\footnote{We say $\alpha$ is the multiplicative error and $\gamma$ is the additive error.} if for every $\e > 0$ there is a monotone, subadditive, convex function $\Psi_\e : \R^d_+ \rightarrow \R$ such that:
     \begin{enumerate} [topsep=0pt]
     		\item \emph{Gradient Stability:} $\nabla \Psi_\e(x + y) \geq \exp(-\epsilon \cdot\|y\| -\delta)\cdot \nabla \Psi_\e(x)$ coordinate-wise for all  $x, y \in  \R^d_+$.
     		\item \emph{Norm Approximation:} $\|x\|\leq \Psi_\e(x) \leq \alpha \|x\| + \frac{\gamma}{\e}$ for all  $x \in  \R^d_+$.
	\end{enumerate}
	(We say that $\Psi_\e$ is a gradient-stable approximation of $\|\cdot\|$ at scale $\e$.)
\end{definition}

	Intuitively, the definition requires the gradient of $\Psi_\e(x)$ not to decrease drastically in any component when adding a small vector $y$ to $x$. For example, the $\ell_1$-norm itself satisfies this property, since it is linear in the non-negative orthant, so its gradient does not change at all. Unfortunately, the vast majority of norms themselves do not fulfill such a property. Consider, for example, the $\ell_\infty$-norm, $\| x \|_\infty := \max_i |x_i|$, which is very ``non-linear'' in the non-negative orthant: For a small $\eta$, its gradient at $(1-\eta, 1)$ is $(0, 1)$ but its gradient at $(1+\eta, 1)$ is $(1, 0)$. However, in this case the standard ``smooth'' approximation of $\ell_\infty$ by the softmax function, defined by $SM_\e(x) := \frac{1}{\e} \ln \big(\sum_i e^{\e x_i}\big)$, is a 0-gradient-stable approximation (see \Cref{sec:lpNorms}). 
	
We note that although there are several notions of  ``smooth'' functions  in the literature, we are not aware of \Cref{def:gradStable} appearing before;  see \Cref{sec:further} for further discussion.

 

	Since gradient-stable functions behave similarly to linear functions, they satisfy many properties that are very useful in the design and analysis of algorithms. In particular, the following two such properties that are central to our $O(\alpha + \gamma)$-approximation results.

\paragraph{Gradient-Stability and a Smooth Game Inequality.}  
Our analyses of greedy algorithms rely on the fact that a gradient-stable function $\Psi_\e$ fulfills the following ``smooth game inequality": Letting $\zero = \Load^{(0)} \leq \ldots \leq \Load^{(T)}$ and $\zero = \Load^{(0)}_* \leq \ldots \leq \Load^{(T)}_*$ be two increasing sequences of vectors, we have
	\begin{equation}
	\label{eq:lambdamusmooth}
		{\textstyle \sum_{t=1}^T} \left(\Psi_\e(\Load^{(t-1)} + y^{(t)}_*) -  \Psi_\e(\Load^{(t-1)})\right) \leq \lambda \Psi_\e(\Load_*^{(T)}) + \mu \big(\Psi_\e(\Load^{(T)}) - \Psi_\e(\zero) \big),
	\end{equation}
	where $y^{(t)}_* = \Load^{(t)}_* - \Load^{(t-1)}_*$ and $\lambda$ and $\mu$ depend on $\epsilon$ and $\delta$ from \Cref{def:gradStable}. To see why this is useful, consider online generalized load balancing and set
 $\Load^{(t)}$ and $\Load^{(t)}_*$ to be the load vectors of our greedy algorithm and the optimum, respectively. By the greedy property,  $\Psi_\e(\Load^{(t-1)} + y^{(t)}) \leq \Psi_\e(\Load^{(t-1)} + y^{(t)}_*)$,  and thus
	\begin{align*}
	    \Psi_\e(\Load^{(T)})-\Psi_\e(\zero) ~&=~ {\textstyle \sum_{t=1}^T } \left(\Psi_\e(\Load^{(t-1)} + y^{(t)}) -  \Psi_\e(\Load^{(t-1)})\right)\\ &~\leq ~ {\textstyle \sum_{t=1}^T } \left(\Psi_\e(\Load^{(t-1)} + y^{(t)}_*) -  \Psi_\e(\Load^{(t-1)})\right)
	    ~\leq~ \lambda \Psi_\e(\Load_*^{(T)}) + \mu \big(\Psi_\e(\Load^{(T)}) - \Psi_\e(\zero) \big).
	\end{align*}
	This implies 
\[
\Psi_\e(\Load^{(T)}) ~\leq~ \frac{\lambda}{1-\mu}\Psi_\e(\Load_*^{(T)}) + \Psi_\e(\zero) .
\]

Note that if we could follow such an argument when $\Psi_\e$ is the actual norm, then we would immediately get an approximation factor of $\frac{\lambda}{1-\mu}$, since $\lVert \zero \rVert = 0$. Unfortunately, \eqref{eq:lambdamusmooth} will usually not hold for norms. As we show, it does hold if $\Psi_\e$ fulfills gradient-stability (see \Cref{lemma:likesmoothgame}). Indeed, for small values of $\epsilon$ and $\delta$, the $\lambda$ and $\mu$ in \eqref{eq:lambdamusmooth} will  get smaller as well. However, this comes at the cost of worse approximation bounds since we also have to control $\Psi_\e(\zero)$ (i.e., the additive error term $\gamma$ in the approximation).

Let us point out that this general argument has similarities to a typical Price of Anarchy analyses in algorithmic game theory. Particularly, Roughgarden~\cite{Roughgarden-JACM15} identified a common approach in these analyses and formalized it by the notion of smooth games, which is similar to \eqref{eq:lambdamusmooth}. Many variants of this notion have been discussed, particularly Thang~\cite{Thang20} also adapted it to analyze greedy algorithms. However, the main difference is that because we construct the function $\Psi_\e$ ourselves, we can control $\lambda$ and $\mu$, in particular ensuring that $\mu < 1$. This comes at the cost of introducing the $\Psi_\e(\zero)$ term, which is necessary because it is not possible to construct gradient-stable approximations of (non-linear) norms with $\Psi_\e(\zero) = 0$.

\paragraph{Gradient-Stability and a Converse to Jensen's Inequality.} Another property that gradient-stability brings is that it allows us to approximate $\Psi_\e$ in terms of its gradients. Note that by convexity, namely Jensen's inequality, we have $\Psi_\e(\Load^{(t-1)} + y^{(t)}) - \Psi_\e(\Load^{(t-1)}) \geq \langle \nabla\Psi_\e(\Load^{(t-1)}), y^{(t)} \rangle$. This way, if we have an increasing sequence $\zero = \Load^{(0)} \leq \ldots \leq \Load^{(T)}$ of vectors, we can lower-bound the increase of $\Psi_\e$ on this sequence by $\Psi_\e(\Load^{(T)}) - \Psi_\e(\zero) \geq \sum_{t=1}^T \langle \nabla\Psi_\e(\Load^{(t-1)}), y^{(t)} \rangle$, where $y^{(t)} = \Load^{(t)} - \Load^{(t-1)}$. Gradient-stability implies an approximate converse of this inequality, namely that $\Psi_\e(\Load^{(T)}) - \Psi_\e(\zero)$ can also be approximately \emph{upper-bounded} in terms of $\sum_{t=1}^T \langle \nabla\Psi_\e(\Load^{(t-1)}), y^{(t)} \rangle$ if all $y^{(t)}$ are bounded (see \Cref{lemma:PsiVsCost}). In our bandit applications, this lets us linearize the problem and use sublinear-regret properties of classic linear bandit algorithms. Again, the bounds are better if $\epsilon$ and $\delta$ are small, which comes at the cost of a worse approximation of underlying the norm.

The biggest challenge in making this notion of gradient-stability truly useful is designing gradient-stable approximations for  norms, which we discuss next.
	

\subsection{Constructing Gradient-Stable  Approximations of Norms}  \label{sec:introGSApprox}

\focs{As pointed out before, most norms themselves do not fulfill gradient-stability. In particular, for every $\ell_p$-norm (except for $p=1$) the partial derivative of the second coordinate decreases from $1$ to almost $0$ even for the small shift from $(0,\eta^2)$ to $(\eta,\eta^2)$ when $\eta$ is tiny.}

Approximating functions so that gradients change less drastically is indeed an important goal in many fields, e.g., in  Online Learning. A standard approach is to add a random perturbation (which is the principle of Follow the Perturbed Leader~\cite{KalaiVemp-Journ05}), i.e.,  define the approximation $\Psi_\e(x) = \E_\nu \|x + \nu\|$ for some random noise vector $\nu$. For example, when each coordinate of $\nu$ is an independent Exponential random variable with mean $1/\e$, standard arguments give that regardless of the starting norm $\|\cdot\|$, we have $\nabla \Psi_\e(u+v) \ge e^{-\e \|v\|_{\infty}} \cdot \nabla \Psi_\e(u)$ for all $u, v \in \R^d_+$. However, such an approximation will  be too weak for our applications. In this case, we get $\Psi_\e(\zero) = \E_\nu \|\nu\| \geq \frac{1}{\e} \| \ones \|$, and so $\gamma \geq \| \ones \|$ in \Cref{def:gradStable} can be as big as $d$, even in case of the $\ell_1$-norm. This makes an $O(\alpha + \gamma)$-approximation in our results very weak, namely it will only be polynomial rather than poly-logarithmic in $d$.
	
	Another natural attempt is to make the noise $\nu$ more adapted to the norm by having it distributed proportional to $e^{-\|x\|}$. In this case, however, we again  get $\Psi_\e(\zero) = d$, which gives an $\Omega(d)$-approximation. (For $\ell_p$-norms actually a ``small'' deterministic noise suffices, but arguing that requires the specific structure of these norms; see \Cref{sec:lpNorms}.)	The other standard technique of adding regularization to the variational definition (i.e.,  $\|x\| = \max_{\|y\|_{\star} \le 1} \ip{y}{x}$  where $\|\cdot\|_{\star}$ is the dual norm) also suffers from the same issue. We discuss other related notions of gradient-stability in \Cref{sec:further}.
	
	Therefore, one of our main technical contribution is to obtain the right notion of gradient-stability for all monotone symmetric norms. For that, we use \mtop{k} norms as our building blocks, since it is a classic fact (at least since Ky Fan's Dominance Theorem~\cite{horn2012matrix}) that they form a sort of ``basis'' for all monotone symmetric norms. Since \mtop{k} includes the $\ell_1$ norm, the random perturbation idea alone will not give gradient-stability as discussed above. In order to achieve gradient-stability for \mtop{k} norms, our main idea is to not only use a random perturbation  but also to \textbf{randomly perturb the parameter $k$}, namely setting $\Psi_\e(x) = \E_{K,\nu} \|x + \nu\|_{\mtop{K}}$. We show that using an appropriately random $K$ with $\E[K] \approx k$, it suffices to use $\nu$ having coordinates Exponentially distributed with mean $\frac{1}{k \e}$ (so a factor $k$ smaller than the traditional one). Then it is easy to show 
	$ \Psi_\e(x) =\E_{K,\nu} \| x + \nu\|_{\mtop{K}} \le O(1) \cdot \| x \|_{\mtop{k}} + \frac{O(\log d)}{\e}$, so that $\alpha = O(1)$ and $\gamma = O(\log d)$. 
	The general idea why this random $K$ helps is as follows: A partial derivative of the $\mtop{K}$-norm is $1$ if and only if this coordinate is among the highest $K$ in the vector, otherwise it is $0$; so, if $K$ is deterministic then it  is possible to increase only one coordinate of the vector to drastically change the partial derivative; however,  randomizing $K$ makes this impossible.
	
	The final step in obtaining a gradient-stable approximation of every symmetric norm is to prove that gradient stability is approximately preserved under compositions. This is useful since we show that any symmetric norm can be approximated (up to an $O(\log d)$ factor) by a composition of \mtop{k} norms. This composition lemma is also of independent interest since it allows us to obtain good approximations for several other online problems where the objective is not symmetric.

\ifx \hasmain \undefined
  \end{document}
\fi

\section{Gradient-Stability for Symmetric Norms  and Compositions}
\label{sec:gradStableSymm}

	The main result of this section shows that every monotone symmetric norm admits a gradient-stable approximation. 
	
	\begin{theorem} \label{thm:symm}
		Consider a monotone symmetric norm $\|\cdot\|$ in $\R^d$. Then for every $\delta \in (0,1)$, 
		there is a $\delta$-gradient-stable  approximation of $\|\cdot\|$ with error 
		\[ \left(2 \cdot e^\delta \cdot ( 1+ \delta^{-1}) \cdot \log d~,~ 4 \big(1+e^\delta \cdot ( 1+ \delta^{-1}) \big) \frac{e^\delta \log^2 d}{\delta}\right).
		\] 
		\focs{Moreover, given Ball-Optimization oracle access to the norm $\|\cdot\|$, we can efficiently perform value and gradient oracle calls to $\delta$-gradient-stable approximation with a slightly larger error of}
		\[ \left(4 \cdot e^\delta \cdot ( 1+ 4 \delta^{-1}) \cdot \log d~,~ 16 \big(1+e^{2\delta} \cdot ( 1+ \delta^{-1}) \big) \frac{e^\delta \log^2 d}{\delta}\right).
		\]		
	\end{theorem}

    For our applications, it suffices to  think of $\delta$ as a small constant like $\delta = \frac{1}{4}$, in which case the error becomes $(O(\log d), O(\log^2 d))$.
    
    En route, we give a generic result (\Cref{lem:composition}) on how composition of norms with gradient-stable approximations also admits a gradient-stable approximation; \focs{not only this result allows us to obtain gradient-stable approximations for many different norms, but may also be useful in future work.}
    
    At a high-level, our approach for proving \Cref{thm:symm} is to: (a) prove in \Cref{sec:topk} that the $\mtop{k}$ norm admits a gradient-stable approximation; (b) prove in \Cref{sec:composition} the aforementioned composition result; (c) show in \Cref{sec:symmNorms} how to use composition of $\mtop{k}$ norms to obtain a gradient-stable approximation for any symmetric norm.
    

\ifx \hasmain \undefined
  \input{preamble}
  \begin{document}
\fi

\subsection{Top-\texorpdfstring{$k$}{k} Norms}  \label{sec:topk}
	
	Recall that for a non-negative vector $x \in \R^d_+$, its $\mtop{k}$ norm $\|x\|_{\mtop{k}}$ is the sum of the $k$ largest coordinates of $x$. We prove the following:
	
	\begin{theorem}\label{thm:topk}
	 	For every $\delta \in (0,1)$, the \mtop{k} norm in $\R^d_+$ admits a $\delta$-gradient-stable approximation with error
	 	\[ \left( \exp(\delta) \Big(1+\frac{1}{\delta}\Big) ~,~ \exp(\delta)\, \frac{H_d}{\delta} \right),
	 	\]
	 	where $H_d = 1 + \frac{1}{2} + \frac{1}{3} + \ldots + \frac{1}{d}$ is the $d$-th harmonic number.
	\end{theorem}

	
	\focs{We will prove this theorem in the remainder of the section and ignore computational questions. In \Cref{sec:topk-polytime} we extend the result to polynomial-time computation using sampling techniques: the resulting approximation is only differentiable almost-everywhere, but this is enough for applications since one can simply add tiny random perturbations to the points where the function is evaluated.
	
	\begin{restatable}{theorem}{thmtopkpolytime}
	\label{thm:topk-polytime}
	 	For every $\delta, \rho \in (0,1)$, there is an algorithm that with probability at least $1 - \rho$ { constructs an approximation of the \mtop{k} norm in $\R^d_+$ with error
	 	\[ \left( \exp(\delta) \Big(1+\frac{4}{\delta} \Big) ~,~ \exp(2 \delta)\, \frac{4 H_d}{\delta} \right)
	 	\] that is differentiable almost-everywhere and wherever it is differentiable it satisfies $\delta$-gradient-stability. }
	 	Moreover, one can perform value and gradient oracle calls in time polynomial in $\frac{d}{\rho}$.
	\end{restatable}
	}
	
	
	\IGNORE{
	\tnote{Or something like}
	\begin{theorem}
	 	For every $\delta \in (0,1)$, $\beta \in (0,1)$, and $\rho \in (0,1)$, there is an algorithm that with probability at least $1 - \rho$ constructs a $\delta$-gradient-stable approximation of the \mtop{k} norm in $\R^d_+$ with error
	 	\[ \left( \exp(\delta) \Big(1+\frac{1 + \beta}{\delta} \Big) ~,~ \exp((1+\beta) \delta)\, \frac{(1+\beta) H_d}{\delta} \right),
	 	\]
	 	where $H_d = 1 + \frac{1}{2} + \frac{1}{3} + \ldots + \frac{1}{d}$ is the $d$-th harmonic number, in time polynomial in $\frac{d}{\beta \rho}$ so that one can perform value and gradient oracle calls in time polynomial in $\frac{d}{\beta \rho}$.
	\end{theorem}
	}

	\paragraph{Construction of a gradient-stable approximation.} We first define the function $f_\e : \R^d \rightarrow \R$, which will be our gradient-stable approximation of the $\mtop{k}$ norm up to a scaling. 
	The function $f_\e : \R^d \rightarrow \R$ is given by 
	\begin{align} \label{eq:topkApprox}
	f_\e(x) := \E_{K,\nu} \|x + \nu\|_{\mtop{K}}, ~\text{ where}
	\end{align}
	\begin{enumerate}
		\item $\nu_i$ is an independent Exponential random variable with rate $\lambda = \epsilon k$ for every $i \in [d]$.
		\item $K$ follows the geometric distribution starting at 1 with parameter $p = 1-q$ for $q = 1-\frac{\delta}{k}$, i.e., $\Pr(K = i) = q^{i-1} (1-q)$ for $i \ge 1$ and $\E[K] = \frac{k}{\delta}$.
		\item For $i > d$, we define $\|x\|_{\mtop{i}} = \|x\|_{\mtop{d}}$.  
	\end{enumerate}
	
	Note that there are \textbf{two perturbations} in $\| x \|_{\mtop{k}}$ to obtain $f_\e(x)$. Firstly, we add a random noise vector $\nu$. This noise is drawn from an exponential distribution with parameter $\lambda$ because we have $\Pr(\nu_i \ge a+b) = e^{-\lambda b} \cdot \Pr(\nu_i \ge a)$ for any non-negative numbers $a$ and $b$. So if $\lambda$ is sufficiently small, the noise $\nu$ helps ``hiding'' any increase in $x$, i.e., it makes $\nabla f_\e(x +y) \approx \nabla f_\e(x)$.
	Such a technique is well-established, e.g., it is used in Follow the Perturbed Leader~\cite{KalaiVemp-Journ05} or differential privacy~\cite{DworkR14}. However, in order to have the desired effect, $\lambda$ would have to be very small, making the approximation very weak, as discussed in Section \ref{sec:introGSApprox}. 
	
	Thus,  we also perturb by considering the $\mtop{K}$-norm for a random $K$, as opposed to the $\mtop{k}$-norm. The $i$-th partial derivative  of $f_\e$ now becomes the probability that the $i$-th component is among the $K$ largest entries in $x + \nu$ (see \eqref{eq:topkGrad1}). If $K = k$ deterministically, it is possible to slightly increase  one entry in $x$ and change this probability by a lot. With a random $K$, this is no longer possible. In order to significantly change the probability, many components in $x$ need to increase.
	
	Indeed,   in \Cref{thm:stab} we will prove that gradients become stable by combining these two perturbations with the respective probability distributions. The idea is as follows: consider any vector $y$  with $\| y \|_{\mtop{k}} \leq 1$ (for simplicity) by which $x$ could be shifted. Then there are most $k$ entries larger than $\frac{1}{k}$. The random choice of $K$ ensures that these large entries do not influence the gradient by too much whereas the random choice of $\nu$ ensures that the effect of the small entries is bounded.

	\paragraph{Guarantees of the function $f_\e$.}	We start by showing that even with the introduction of the two perturbations, $f_\e$ still gives a good approximation to the $\mtop{k}$ norm. 
	
 
 In the rest of the section, we denote the function $f_\e(\cdot)$ by $f(\cdot)$ for ease of notation.
	
	\begin{lemma} \label{lem:approxTopk} 	 The function $f$ defined in \Cref{eq:topkApprox} satisfies
	$$f(x) \ge \exp(-\delta) \cdot \|x\|_{\mtop{k}} \quad \text{and} \quad f(x) \le \left( 1 + \frac{1}{\delta} \right) \cdot \|x\|_{\mtop{k}} + \frac{H_d}{\delta \epsilon}.$$
	Moreover, $f$ is monotone, subadditive, and convex. 
	\end{lemma}
	\begin{proof}
	We first show the lower bound. By the choice of the distributions for $K$ and $\nu$, we have
	\begin{align*}
	f(x) ~& =~ \E_{K,\nu} \|x + \nu\|_{\mtop{K}} ~\geq~ \E_{K} \|x\|_{\mtop{K}} ~=~ \sum_{j=1}^\infty \Pr(K=j) \|x\|_{\mtop{j}} \\
	~& \geq~ \Pr(K \geq k) \|x\|_{\mtop{k}} ~=~ q^{k-1} \|x\|_{\mtop{k}} ~=~ \left(1 - \frac{\delta}{k}\right)^{k-1} \|x\|_{\mtop{k}} ~\geq~ \exp(-\delta) \|x\|_{\mtop{k}}.
	\end{align*}
	
	To obtain the upper bound, we can use that the $\mtop{K}$-norm satisfies the triangle inequality
	\[
	f(x) ~=~ \E_{K,\nu} \|x + \nu\|_{\mtop{K}} ~\leq~ \E_{K,\nu} \big(\|x\|_{\mtop{K}} + \|\nu\|_{\mtop{K}} \big) ~\leq~ \E_{K,\nu}\left[\|x\|_{\mtop{K}} + K \max_i \nu_i\right].
	\]
	We can upper-bound the two terms on the right-hand side by 
	\begin{align*}
	\E_K \|x\|_{\mtop{K}} \,\leq\, \E_K \Big[ \max\left\{1, \tfrac{K}{k}\right\} \cdot \|x\|_{\mtop{k}}\Big] \,\leq\, \|x\|_{\mtop{k}} \left(1 + \frac{1}{k} \E [K] \right) \,=\, \|x\|_{\mtop{k}} \left( 1 + \frac{1}{\delta} \right) 
	\end{align*}
	and $\E_{K,\nu} \big[ K \max_i \nu_i \big] = \frac{k}{\delta} \frac{H_d}{\lambda} = \frac{H_d}{\delta \epsilon}$ since the expected maximum of $d$ exponentials with rate $\lambda$ equals $\frac{H_d}{\lambda}$ \cite[Page 73, Equation (4.6.6)]{doi:10.1137/1.9780898719062}.
	
	The monotonicity, subadditivity, and convexity of $f$ follows from the fact that the $\mtop{k}$ norms satisfy these properties and that $f$ is formed by taking convex combinations of these norms.
	\end{proof}
	
    Now we have the main technical lemma, which proves the gradient-stability of $f$.
	
	\begin{lemma} \label{thm:stab}
		For all $i$ and all non-negative $x$ and $y$, we have 
		\[ \nabla_i f(x+y) ~\ge~ \exp(-\epsilon \|y\|_{\mtop{k}}-\delta) \cdot \nabla_i f(x).
		\]
	\end{lemma}

\begin{proof}
	Due to symmetry, it suffices to prove this claim for the first coordinate without loss of generality, i.e.,  $i=1$. Given a vector $x$, we use $x_{(j)}$ to denote its $j$-th largest coordinate for $j\in \{1,\ldots, d\}$ and to denote $x_{(j)}=0$ for $j>d$. We also use $x_{-1} \in \R^{d-1}$ to denote the vector $x$ without its first coordinate.
	
	Notice that the $\mtop{j}$ norm is differentiable at any point $x$ with pairwise distinct coordinates and that in this case we have $\nabla_1 \|x\|_{\mtop{j}} = 1$ if the first coordinate is among the highest $j$ coordinates of $x$, or equivalently, $x_1 > (x_{-1})_{(j)}$, and otherwise $\nabla_1 \|x\|_{\mtop{j}} = 0$. Note that the coordinates of $x+\nu$ will be pairwise distinct almost surely. So, by linearity of expectation, this implies
	\begin{align}
		\nabla_1 f(x) \,&=\, {\textstyle \sum_{j=1}^\infty}  \Pr(K=j)\cdot \E_\nu \Big[\nabla_1 \|x + \nu\|_{\mtop{j}}\Big]	\notag\\
		\,&=\, {\textstyle \sum_{j=1}^\infty} \Pr(K=j)\cdot \Pr\Big(x_1 + \nu_1 > ((x+\nu)_{-1})_{(j)}\Big),  \label{eq:topkGrad1}
	\end{align}
	and analogously,
	\[
		\nabla_1 f(x+y) ~=~ {\textstyle \sum_{j=1}^\infty} \Pr(K=j)\cdot \Pr\Big(x_1 + y_1 + \nu_1 > ((x+y+\nu)_{-1})_{(j)} \Big).
	\]
	
	Recall that we want to show that $\nabla_1 f(x+y)$ is not much smaller than $\nabla_1 f(x)$. Comparing the two sums term by term will not be successful because for a fixed $j$ increasing only a single entry in $x$ might drastically decrease the probability. Instead, we compare the $j$-th term in the first sum to the $(j+k-1)$-th term in the second sum. By our choice of $K$ being drawn from a geometric distribution with parameter $q$, we have $\Pr(K=j+k-1) = q^{k-1} \Pr(K=j)$, and thus 
	\begin{align}
		\nabla_1 f(x + y)	~&\geq~ {\textstyle \sum_{j = k}^\infty } \Pr(K=j)\cdot \Pr\Big(x_1 + y_1 + \nu_1 > ((x+y+\nu)_{-1})_{(j)}\Big)  \notag\\
		&=~ {\textstyle \sum_{j = 1}^\infty } \Pr(K=j+k-1)\cdot \Pr\Big(x_1 + y_1 + \nu_1 > ((x+y+\nu)_{-1})_{(j+k-1)}\Big)  \notag\\
		&= q^{k-1} \cdot {\textstyle \sum_{j = 1}^\infty } \Pr(K=j)\cdot \Pr\Big(x_1 + y_1 + \nu_1 > ((x+y+\nu)_{-1})_{(j+k-1)}\Big).   \label{eq:topkGrad2}
	\end{align}	

	To conclude the comparison with the right-hand side of \eqref{eq:topkGrad1}, we bound the probability involving $((x+y+\nu)_{-1})_{(j+k-1)}$ in terms of that involving $((x+\nu)_{-1})_{(j)}$ using the following claim.

	\begin{claim} \label{lemma:gamma}
		For every vector $u \in \reals_+^d$  and every vector $v \in \reals_{+}^d$, we have for all $j \geq 1$ that $
		(u+v)_{(j+k-1)} \le u_{(j)} + v_{(k)}  \le  u_{(j)} +  \frac{\|v\|_{\mtop{k}}}{k}.
		$
	\end{claim}
	
	\begin{proof}
        We first observe that there are at most than $j-1$ coordinates $i \in [d]$ for which $u_i > u_{(j)}$.
	    Next, we  observe that there are at most $k-1$ coordinates $i \in [d]$ such that $(u+v)_i > u_{(j)} + v_{(k)}$ but $u_i \leq u_{(j)}$. To see this, suppose otherwise. Then there is a set $S \subseteq [d]$ of size $k$ such that $(u+v)_i > u_{(j)} + v_{(k)}$ and $u_i \leq u_{(j)}$ for all $i \in S$. This implies that $v_i > v_{(k)}$ for every $i \in S$, which is a contradiction.
	    
	    Combining these two observations, there are no more than $k+j-2$ coordinates $i \in [d]$ for which $(u+v)_i > u_{(j)} + v_{(k)}$, meaning that $(u+v)_{(j+k-1)} \leq u_{(j)} + v_{(k)}$.
	\end{proof}
	
	Now \Cref{lemma:gamma} implies that
	\begin{align*}
	\Pr\Big(x_1 + y_1 + \nu_1 > \big((x+y+\nu)_{-1}\big)_{(j+k-1)}\Big) & ~\geq~ \Pr\Big(x_1 + y_1 + \nu_1 \,>\, \big((x+\nu)_{-1}\big)_{(j)} + \tfrac{\|y\|_{\mtop{k}}}{k}\Big) \\
	& ~\geq~ \Pr\Big(\nu_1 \,>\, \big((x+\nu)_{-1}\big)_{(j)} + \tfrac{\|y\|_{\mtop{k}}}{k} - x_1\Big).
	\end{align*}
	Since $\nu_1$ is drawn independently from an exponential distribution with rate $\lambda$,  for every pair of non-negative numbers  $a,b$ we have $\Pr(\nu_1 \ge a+b) = e^{-\lambda b} \cdot \Pr(\nu_1 \ge a)$. Therefore, the previous inequality becomes
	\begin{align*} 
	\Pr\Big(x_1 + y_1 + \nu_1 > \big((x+y+\nu)_{-1}\big)_{(j+k-1)}\Big) & \geq \exp\Big(- \tfrac{\lambda \|y\|_{\mtop{k}}}{k}\Big) \Pr\Big(\nu_1 > \big((x+\nu)_{-1}\big)_{(j)} - x_1\Big) \notag \\
	& = \exp\Big(- \e \|y\|_{\mtop{k}} \Big) \Pr\Big(x_1 + \nu_1 > \big((x+\nu)_{-1}\big)_{(j)}\Big),
	\end{align*}
	where the last equation uses the fact that we set $\lambda = \e k$.
	
	With this at hand, we can employ it on \eqref{eq:topkGrad2} to obtain the gradient lower bound 
	\begin{align*}
		\nabla_1 f(x + y) ~&\geq~ q^{k-1} \cdot \exp\Big(- \e  \|y\|_{\mtop{k}}\Big)\cdot {\textstyle \sum_{j = 1}^\infty}  \Pr(K=j)\cdot \Pr\Big(x_1 + \nu_1 > ((x+\nu)_{-1})_{(j)}\Big)\\
		&=~ q^{k-1} \cdot \exp\Big(- \e \|y\|_{\mtop{k}}\Big) \cdot \nabla_1 f(x).
	\end{align*}
	But by our choice of $q$, we have $q^{k-1} = (1 - \frac{\delta}{k})^{k-1} \geq \exp(-\delta)$, which  concludes the proof of \Cref{thm:stab}. 
\end{proof}

	\paragraph{Proof of Theorem \ref{thm:topk}.}
We can now construct a gradient-stable approximation of $\mtop{k}$.	We claim that the function $\Psi_\e$ given by $\Psi_\e(\cdot) = \exp(\delta) f_\e(\cdot)$ is a $\delta$-gradient-stable approximation of the $\mtop{k}$ norm at scale $\e$ with the desired error $(\exp(\delta)\,(1+\frac{1}{\delta}) \,,\, \exp(\delta) \frac{H_d}{\delta})$. To see this, first from \Cref{lem:approxTopk} we see that $\Psi_\e$ is monotone,  subadditive, and convex. Also from this lemma we get the error bound 
	\begin{align*}
	\|\cdot\|_{\mtop{k}} \,\le\, \Psi_\e(\cdot) \,\le\, \exp(\delta) \bigg(1 + \frac{1}{\delta}\bigg) \, \|\cdot\|_{\mtop{k}} +  \frac{\exp(\delta) H_d}{\delta \e}.
	\end{align*}
	Finally, from \Cref{thm:stab} and linearity of the gradient operator, we directly get 
	\begin{align*}
		\nabla_i \Psi_\e(x+y) \,\ge\, \exp(-\epsilon \|y\|_{\mtop{k}}-\delta) \cdot \nabla_i \Psi_\e(x),~~~~~~\forall x,y \in \R^d_+,
	\end{align*}
	and so $\Psi_\e$ has all the claimed properties of gradient stability.
	
	\IGNORE{\paragraph{Computation.} The value and gradient of function $f(\cdot)$ are given by \eqref{eq:topkApprox} and \eqref{eq:topkGrad1}. Since $K$ has support in $\{0,\ldots, d\}$, it suffices to be able to compute these expressions for a given $K=j$. In \eqref{eq:topkApprox} we need to compute $ \E_{\nu} \|x + \nu\|_{\mtop{j}}$ and in \eqref{eq:topkGrad1} we need to compute $\Pr\big(x_1 + \nu_1 > ((x+\nu)_{-1})_{(j)}\big)$, where $\nu_i$ is an independent Exponential random variable with rate $\lambda = \epsilon k$ for every $i \in [d]$. ....
	}

	Since we can obtain such an approximation for every scale $\e > 0$, this proves that the $\mtop{k}$ norm admits the desired gradient-stable approximation. This concludes the proof of the theorem.

\ifx \hasmain \undefined
  \end{document}
\fi


\ifx \hasmain \undefined
  \input{preamble}
  \begin{document}
\fi

\renewcommand{\bar}{\tilde}

\subsection{Composition of Norms} \label{sec:composition}

	

	We now show that gradient-stability is approximately preserved when considering a ``nested'' composition of norms. We remark that we assume the normalization that the outer norm $\|\cdot\|_{(0)}$ is lower bounded by $\|\cdot\|_{\infty}$ only to obtain a cleaner statement, since this can always be satisfied by an appropriate scaling of the norm. 

	\begin{restatable}[Norm Composition]{theorem}{compNorms} \label{lem:composition}
		Consider monotone (not necessarily symmetric) ``inner'' norms $\|\cdot\|_{(1)}, \ldots, \|\cdot\|_{(\ell)}$ over $\R^d$, and non-negative matrices $A_i \in \R_+^{d \times n}$. Also consider a monotone ``outer'' norm $\|\cdot\|_{(0)}$ over $\R^\ell$, and assume that $\|\cdot\|_{\infty} \le \|\cdot\|_{(0)}$. Then consider the composed norm over $\R^n$:
		\begin{align*}
			\norm(x) := \bigg\|\Big(\|A_1 x\|_{(1)},~\ldots~,\|A_\ell x\|_{(\ell)} \Big) \bigg\|_{(0)}.
		\end{align*}
		
		If each norm $\|\cdot\|_{(i)}$ admits a $\delta_i$-gradient-stable approximation with error $(\alpha_i, \gamma_i)$, then for every $c > 0$, $\norm(\cdot)$ admits a $\delta$-gradient-stable approximation with error $(\alpha, \gamma)$, where:
		\begin{align*}
			\alpha = \alpha_0 \cdot \big(\max_{i \ge 1} \alpha_i\big),~~~~~~ \delta = \delta_0 + \max_{i \ge 1} \delta_i,~~~~~~ \gamma = \big(\max_{i \ge 1} \alpha_i + c\big) \cdot \bigg[\frac{\alpha_0}{c} \|(\gamma_1,\ldots,\gamma_\ell)\|_{(0)} + \gamma_0 \bigg].
		\end{align*}
	    \focs{Moreover, given value and gradient oracles to the $\delta_i$-gradient-stable approximations of inner and outer norms, we can efficiently simulate value and gradient oracles to this $\delta$-gradient-stable approximation.}
	\end{restatable}
	
	Its proof is presented in \Cref{app:normComp}. In a typical use the $\alpha_i$'s and $\delta_i$'s are constants, $c = 1$, and the outer norm equals $\|\cdot\|_{\infty}$, in which case we get $\alpha = \delta = O(1)$ and $\gamma = O(\max_{i \ge 0} \gamma_i)$; however, the extra flexibility of the parameters will be needed in some of our applications.

    As a corollary of the previous lemma, we observe that gradient-stability is invariant with respect to scaling the norm, a fact that will be useful later. This is obtained by applying \Cref{lem:composition} with a single norm $\|\cdot\|_{(1)} = \|\cdot\|$ ($\ell=1$), with $A_1$ being the diagonal matrix with $\beta$'s in the diagonal, $\|\cdot\|_{(0)} = \|\cdot\|_1$ (recall that the latter admits a $0$-gradient-stable approximation with error $(1,0)$, see \Cref{lemma:smoothLp}), and $c \rightarrow \infty$.
	
	\begin{lemma}[Norm Scaling] \label{lem:scaling}
		If the norm $\|\cdot\|$ admit a $\delta$-gradient-stable approximation with error $(\alpha, \gamma)$, then so does  the scaled norm $\beta \|\cdot\|$ for every $\beta > 0$ (with the same parameters).
	\end{lemma}	
	

\ifx \hasmain \undefined
  \end{document}
\fi


\ifx \hasmain \undefined
  \input{preamble}
  \begin{document}
\fi

\subsection{Symmetric Norms} \label{sec:symmNorms}


    Given the smoothing for $\mtop{k}$ norms and the composition property above, we now show that every symmetric norm $\|\cdot\|$ can be approximated by a norm $\vertm{\cdot}$ that is the composition of $\mtop{k}$ norms. \focs{Furthermore, this approximation can be computed in polynomial time given \ballopt oracle access to $\|\cdot\|$.}
    
		

	
	The starting point is that any symmetric monotone norm can be expressed as the supremum of positive combinations of \mtop{k} norms. The following follows from Theorem~1.3 of~\cite{kyFanDominance} or Lemma~5.2 of~\cite{CS-STOC19}.

	\begin{lemma} \label{lem:symmToTopkComb}
		For any symmetric monotone norm $\|\cdot\|$ in $\R^d$, there is a set $\W$ of non-negative weights $w \in \R^d_+$ such that for all $x$, we have
		\begin{align}
		\|x\| = \max_{w \in \W} { \sum_{i=1}^d } w_i \|x\|_{\mtop{i}}. \label{eq:symmTop}
		\end{align}
	\end{lemma}
	
	

	While this it not exactly in the format of the composition result from \Cref{lem:composition}, it can be used to obtain an approximation of $\|\cdot\|$ that does have this property.

	\begin{lemma} \label{lem:strucSymm}
		For any symmetric monotone norm $\|\cdot\|$ in $\R^d$, there are $\log d$ non-negative scalars $c_1,c_2\ldots,c_{\log d}$ such that the norm 
		\begin{align}
		\vertm{x} := \bigg\|\Big(c_1 \|x\|_{\mtop{2^1}},~\ldots~, c_{\log d} \|x\|_{\mtop{2^{\log d}}}\Big)  \bigg\|_{\infty} \label{eq:strucSymm}
		\end{align}
		satisfies $\|x\| \,\le\, \vertm{x} \,\le\, 2 \log d \cdot \|x\|.$
	\end{lemma}
	
	\focs{Note that a similar property has been proved in \cite{CS-STOC19}. See \Cref{rem:InftyVsOne} for a discussion.}
	
	\begin{proof}
		Let $\W$ be the set given by the previous lemma applied to the norm $\|\cdot\|$. First, using a bucketing strategy,  we sparsify the sums appearing in \eqref{eq:symmTop} so that each has $\log d$ terms, losing a factor of 2. 
		
		More precisely, consider a $w \in \W$, and the sum $\sum_{i=1}^d w_i \|x\|_{\mtop{i}}$. Round \textbf{up} each $i$ to the nearest power of 2, let $B_j$ be all the $i$'s that get rounded to the power $2^j$. Notice that when we round $i$ to the nearest power, say $i'$, then the $\mtop{i}$ vs $\mtop{i'}$ changes by at most a factor of 2. This gives that 
		\begin{align*}
		\sum_{i=1}^d w_i \|x\|_{\mtop{i}} ~\le~ \sum_{j = 1}^{\log d} \Big(\sum_{i \in B_j} w_i\Big)\,\|x\|_{\mtop{2^j}} \quad \text{and} \quad 
		\sum_{i=1}^d w_i \|x\|_{\mtop{i}} ~\ge~ \frac{1}{2} \sum_{j = 1}^{\log d} \Big(\sum_{i \in B_j} w_i\Big)\,\|x\|_{\mtop{2^j}} \enspace .
		\end{align*}
	Then define $w'$ as $w'_j := \sum_{i \in B_j} w_i$, and add it to the set $\W'$. 
	
	Repeating this operation for all $w \in \W$ gives a set $\W' \subseteq \R^{\log d}_+$ that provides the desired ``sparse'' approximation to $\|\cdot\|$ (a similar inequality has been used in \cite{CS-STOC19}):
	\begin{align*}
		\|x\| ~\le~ \max_{w' \in \W'}  \sum_{j \le \log d} w'_j  \|x\|_{\mtop{2^j}} ~\le~ 2 \|x\| \enspace .
	\end{align*}
	Moreover, since each sum in the previous expression has only $\log d$ terms, replacing each sum by a max only loses a $\log d$ factor, namely:
		\begin{align*}
			\frac{1}{\log d}\cdot \|x\| ~\le~\max_{w'\in \W'} \max_{j \le \log d} w'_j \|x\|_{\mtop{2^j}} ~\le~ 2 \|x\| \enspace .
		\end{align*}
	Setting $c_j := \log d \cdot \max_{w'\in \W'} w'_j$ gives the desired result. 
	\end{proof}
	

    With all these elements, we can finally prove that every monotone symmetric norm admits a gradient-stable approximation. 
    
	
	\begin{proof}[Proof of \Cref{thm:symm}]	
	From \eqref{eq:strucSymm} we see that $\vertm{\cdot}$ is a composition of norms covered by \Cref{lem:composition} with the matrices $A_i$ being the identity; thus, we will employ it to obtain the desired gradient-stable approximation of $\vertm{\cdot}$ (and then of the original norm $\|\cdot\|$). 

	More precisely, fix $\delta \in (0,1)$ and let $\bar{\alpha} = e^\delta ( 1+ \frac{1}{\delta})$ and $\bar{\gamma} = \exp(\delta) \frac{H_d}{\delta}$ be the parameters from \Cref{thm:topk}. Combining the gradient-stable approximation of $\mtop{k}$ norms (\Cref{thm:topk}) and the invariance with respect to scaling the norms (\Cref{lem:scaling}), we get that the scaled norm $c_j \|\cdot\|_{\mtop{2^j}}$ that appears in $\vertm{\cdot}$ admits a $\delta$-gradient-stable approximation with error $(\bar{\alpha}, \bar{\gamma})$. Also recall that the norm $\|\cdot\|_{\infty}$ with $(\log d)$-coordinates admits a $0$-gradient-stable approximation with error $(1,\ln \log d)$  (\Cref{lem:SM}). Thus, applying composition \Cref{lem:composition} with (parameter $c= 1$) we get that $\vertm{\cdot}$ admits a $\delta$-gradient-stable approximation with error $(\alpha, \gamma)$, where $\alpha = \bar{\alpha}$ and 
	\[ 
	\gamma ~=~ (\bar{\alpha} + 1) \cdot \bigg[ \bar{\gamma} + \ln \log d\bigg] ~\le~ 2 (\bar{\alpha} + 1) \cdot \frac{e^{\delta} \log d}{\delta}.
	\]
	
	We claim that this gives the desired gradient-stable approximation of the original norm $\|\cdot\|$. For that, let $h$ be a $\delta$-gradient-stable approximation of $\vertm{\cdot}$ with error $(\alpha, \gamma)$ at scale $\e$. Then this is a $\delta$-gradient-stable approximation of $\|\cdot\|$ with error $(2 \alpha \log d, 2 \gamma \log d)$ at scale $\e' = 2 \e \log d$: the error is 
	\begin{align*}
		\|\cdot\| ~\le~ \vertm{\cdot} ~\le~ h(\cdot) ~\le~  \alpha \vertm{\cdot} + \frac{\gamma}{\e} ~\le~ 2 \alpha (\log d) \|\cdot\| + \frac{2 \gamma \log d}{\e'},
	\end{align*} 
	and the gradient stability
	\begin{align*}
		\nabla h(u+v) \ge e^{-\e \vertm{v} - \delta} \cdot  \nabla h(u) \ge e^{-\e' \|v\| - \delta} \cdot \nabla h(u),~~~~~\forall u,v \in \R^d_+.
	\end{align*} 	
	Since this construction holds for every $\e' > 0$, the norm $\|\cdot\|$ admits the desired approximation. Unpacking the values of $\alpha$ and $\gamma$ concludes the \focs{existential part of the} proof of \Cref{thm:symm}.
	
	\focs{\paragraph{Computation.} Note that if for a symmetric norm $\|\cdot\|$ we are explicitly given $\W$ in  \eqref{eq:symmTop}, then in time polynomial in $d$ and $\lvert \W \rvert$ we can obtain the $\delta$-gradient stable approximation $h$ of $\vertm{\cdot}$ with error $(\alpha, \gamma)$ at scale $\e$ using the computational part of \Cref{thm:topk} and \Cref{lem:composition}. If instead, we are only given \ballopt oracle access to $\|\cdot\|$, we can reduce the problem to one with a poly-size $\W$ using the following lemma.
	
	\begin{restatable}{lemma}{SymmetricNormExplicit}
		For any symmetric monotone norm $\|\cdot\|$ in $\R^d$ and every $\epsilon' > 0$, there is a set $\W'$ of non-negative weights $w \in \R^d_+$ with $\lvert \W' \rvert \leq \left(\frac{d}{\epsilon'} \right)^{O(1/\epsilon')}$ that can be computed using  \ballopt oracle access to  $\|\cdot\|$ in time $\left(\frac{d}{\epsilon'} \right)^{O(1/\epsilon')}$ such that for all $x \in \R_+^d$ we have
		\begin{align}
		(1 - \epsilon') \|x\| ~\leq ~ \max_{w \in \W'} { \sum_{i=1}^d } w_i \|x\|_{\mtop{i}} ~\leq~ (1 + \epsilon') \|x\|.
		\end{align}
	\end{restatable}
	
	This lemma has essentially been shown in \cite{CS-STOC19} as Theorem~5.4. For completeness, we provide the proof in \Cref{sec:SymmetricNormExplicit} in a simplified version with an improved bound.
	
	Using this lemma, we obtain the computational result of \Cref{thm:symm} by first computing the set $\W'$ and then applying the above approximation using \Cref{thm:topk-polytime} on the norm $f$ defined by $f(x) = \max_{w \in \W'} { \sum_{i=1}^d } w_i \|x\|_{\mtop{i}}$. In combination, we only lose a constant factor compared to the existential result.
	}
	\end{proof}

	\focs{
	\begin{remark}\label{rem:InftyVsOne} Another approach to prove  \Cref{thm:symm} is to use a different construction  for the  $\log d$-approx $\vertm{\cdot}$ of $\|\cdot\|$ in \Cref{lem:strucSymm} where the outer $\ell_\infty$ norm  is replaced with an $\ell_1$ norm. (Such an $\ell_1$ approx has been previously given in  \cite{CS-STOC19}.) Although this approach has the advantage that the proof of the composition theorem for $\ell_1$ is much easier than the general \Cref{lem:composition}, it loses an extra $\log d$ factor in the competitive ratio since the $\gamma$ parameter adds up for $\ell_1$. Moreover, our general composition theorem will play a crucial role in some of our applications later where the outer norm is not $\ell_1$.
	\end{remark}
	}

\ifx \hasmain \undefined
  \end{document}
\fi

\section{Gradient-Stability  implies a Smooth Game Inequality}

In this section we prove a smooth game inequality for gradient-stable norm approximations. As discussed in \Cref{sec:introSGI}, this will play a central role in our design of both online and bandit algorithms.
Before showing the inequality, we prove another important ``approximate supermodularity'' property satisfied by gradient-stable approximations. 
 
\begin{lemma}[Approximate Supermodularity] \label{lem:approxSuper}
If $\Psi_\e$ is a $\delta$-gradient-stable approximation of a norm $\|\cdot \|$ at scale $\e$, then for any vectors $\Load, y,z \in \reals_{+}^d$ it satisfies
\[  \Psi_\e(\Load+y+z) - \Psi_\e(\Load+z) ~\geq~  \exp(-\epsilon \cdot\|z\| -\delta)  \cdot \Big( \Psi_\e(\Load+y) - \Psi_\e(\Load) \Big) . \]
\end{lemma}

\begin{proof}
Using gradient-stability and non-negativity of $\nabla \Psi_\e$, we have that
\begin{align*}
     \Psi_\e(\Load+y+z) - \Psi_\e(\Load+z) ~&=~ \int_0^1   \langle \nabla \Psi_\e(\Load+z+\tau y) \,,\, y \rangle ~ d\tau \\
     ~&\geq~    \exp(-\epsilon \cdot\|z\| -\delta)  \cdot \int_0^1   \langle \nabla \Psi_\e(\Load+\tau y) \,,\, y \rangle~ d\tau \\
     ~&=~  \exp(-\epsilon \cdot\|z\| -\delta)  \cdot \Big( \Psi_\e(\Load+y) - \Psi_\e(\Load) \Big) .   \qedhere
\end{align*}
\end{proof}

We use approximate supermodularity to prove the following smoothness.

\begin{lemma}[Smooth Game Inequality]
\label{lemma:likesmoothgame}
Let $\zero = \Load^{(0)} \leq \Load^{(1)} \leq  \ldots \leq \Load^{(T)}$ and $\zero = \Load_*^{(0)} \leq \Load_*^{(1)} \leq  \ldots \leq \Load_*^{(T)}$ be any two non-decreasing sequences of non-negative vectors and let $y^{(t)} = \Load^{(t)} - \Load^{(t-1)}$ and $y^{(t)}_* = \Load^{(t)}_* - \Load^{(t-1)}_*$. Then we have
\[
\sum_{t=1}^T \left(\Psi_\e(\Load^{(t-1)} + y^{(t)}_*) -  \Psi_\e(\Load^{(t-1)})\right) \leq \zeta \Psi_\e(\Load_*^{(T)}) + \left(\zeta - 1\right) \left(\Psi_\e(\Load^{(T)}) - \Psi_\e(\Load^{(0)}) \right),
\]
where $\zeta = \exp\left(\epsilon \lVert \Load_*^{(T)} \rVert +\delta\right)$. 
\end{lemma}


\begin{proof}
The idea is to consider the sequence $\Load^{(t)} + \Load_*^{(t)}$ in which we add both $y^{(t)}$ and $y^{(t)}_*$ in every step. The  increases of $\Psi_\e$ can then be related using approximate supermodularity and subadditivity.

We  write
\begin{align*}
 \Psi_\e(\Load^{(t)} + \Load_*^{(t)} ) - \Psi_\e(\Load^{(t-1)} + \Load_*^{(t-1)}) 
~& =~ \left(\Psi_\e(\Load^{(t)} + \Load_*^{(t)}) - \Psi_\e(\Load^{(t-1)} + \Load_*^{(t)} ) \right) \\
& \qquad + \left(\Psi_\e(\Load^{(t-1)} + \Load_*^{(t)} ) - \Psi_\e(\Load^{(t-1)} + \Load_*^{(t-1)}) \right).
\end{align*}
Furthermore, by approximate supermodularity property in \Cref{lem:approxSuper}, we have
\begin{align*}
 \Psi_\e(\Load^{(t)} + \Load_*^{(t)}) - \Psi_\e(\Load^{(t-1)} + \Load_*^{(t)} )  \geq \exp\left(-\epsilon \lVert \Load_*^{(t)}  \rVert -\delta\right) \left( \Psi_\e(\Load^{(t)}) - \Psi_\e(\Load^{(t-1)}) \right)
\end{align*}
and
\begin{align*}
 \Psi_\e(\Load^{(t-1)} + \Load_*^{(t)} ) - \Psi_\e(\Load^{(t-1)} + \Load_*^{(t-1)}) 
 \geq \exp\left(-\epsilon \lVert \Load_*^{(t-1)} \rVert -\delta\right) \left( \Psi_\e(\Load^{(t-1)} + y^{(t)}_*) - \Psi_\e(\Load^{(t-1)}) \right).
\end{align*}

By monotonicity of the norm, we have $\lVert \Load_*^{(t-1)} \rVert \leq \lVert \Load_*^{(T)} \rVert$ as well as $\lVert \Load_*^{(t)} \rVert \leq \lVert \Load_*^{(T)} \rVert$. So, we get
\begin{align*}
 \Psi_\e(\Load^{(t)} + \Load_*^{(t)}) - \Psi_\e(\Load^{(t-1)} + \Load_*^{(t-1)}) 
 \geq \frac{1}{\zeta} \left( \Psi_\e(\Load^{(t-1)} + y^{(t)}_*) - \Psi_\e(\Load^{(t-1)}) + \Psi_\e(\Load^{(t)} ) - \Psi_\e(\Load^{(t-1)}) \right) .
\end{align*}
Taking the sum over all $t$, two of the sums telescope, so we get
\begin{align*}
 \Psi_\e(\Load^{(T)} + \Load_*^{(T)}) - \Psi_\e(\Load^{(0)})  ~\geq ~ \frac{1}{\zeta} \Big( \sum_{t=1}^T \left( \Psi_\e(\Load^{(t-1)} + y^{(t)}_*) - \Psi_\e(\Load^{(t-1)}) \right) + \Psi_\e(\Load^{(T)}) - \Psi_\e(\Load^{(0)}) \Big) .
\end{align*}

By subadditivity of $\Psi_\e$, we furthermore have $\Psi_\e(\Load^{(T)} + \Load_*^{(T)}) \leq \Psi_\e(\Load^{(T)}) + \Psi_\e(\Load_*^{(T)})$. In combination, this proves the lemma.
\end{proof}

\section{Applications to Online Algorithms}
\ifx \hasmain \undefined
  \input{preamble}
  \begin{document}
\fi

In this section we  use the smooth game inequality  to analyze our online algorithms.

\subsection{Online Generalized Load Balancing}

Recall the \emph{Online Generalized Load Balancing} problem (\loadBal) defined in the introduction.  There are $m$ machines, and $T$ jobs come one-by-one. In each time step $t \in [T]$  a cost matrix $C^{(t)} \in \reals_+^{m \times k}$ comes, indicating the $k$ processing options for this job. After seeing $C^{(t)}$ the goal is to immediately make a scheduling decision $x^{(t)} \in \{e_1, \ldots, e_k\}$ and incur a load vector $ C^{(t)} x^{(t)}$. To measure the quality of the solution, it is given a norm $\|\cdot\|$ over $\R^m$, and the goal is to minimize the norm of the total load vector $\Load^{(T)}:= \sum_{t=1}^T C^{(t)} x^{(t)}$. Let $\OPT$ denote the cost of the optimal offline solution. 

	The following gives a general result for this problem.

\begin{theorem} \label{thm:loadBal}
	Consider the problem \loadBal with a norm $\|\cdot\|$. Fix $\delta=\frac{1}{4}$, and suppose $\|\cdot\|$ has gradient-stable approximations with uniform error $(\alpha,\gamma)$. Then there exists an $O(\alpha + \gamma)$-competitive algorithm for this problem. \focs{Moreover, this algorithm is efficient given value oracle access to the gradient-stable approximation.}
 \end{theorem}

	Together with the gradient-stable approximation of the $\ell_p$ norms from \Cref{lemma:smoothLp}, this gives an $O(\min\{p, \log m\})$ competitive ratio in this special case. Instead, using the approximation of general monotone symmetric norms from \Cref{thm:symm} gives an $O(\log^2 m)$ competitive ratio in this general case, which proves \Cref{thm:loadBalSymm}.	

	\begin{proof}[Proof of \Cref{thm:loadBal}]
	The algorithm that achieves the guarantee is the following: First, by the standard guess-and-double trick we assume without loss of generality that we have an estimate $\widehat{\OPT}$ within a factor of two of the offline optimum $\OPT$, namely $\widehat{\OPT} \in [\OPT, 2\OPT]$~\cite{aspnes}. Given this estimate, the algorithm takes a $\delta$-gradient-stable approximation $\Psi=\Psi_\e$ of $\|\cdot\|$ with parameters $\e = \frac{1}{4 \widehat{\OPT}}$ and $\delta = \frac{1}{4}$ and make scheduling decisions greedily with respect to $\Psi$. 
	
	\focs{Observe that this algorithm requires only $k$ calls to the value oracle of $\Psi$ per time step.}
	
		\begin{algorithm}[H]
		\caption{Gradient-Stable Greedy}
		\begin{algorithmic}[0]      
				\vspace{-2pt}
				\State \textbf{input:} Estimate $\widehat{\OPT} \in [\OPT, 2\,\OPT]$
				\vspace{2pt}
				\State Let   $\Psi=\Psi_\e$ be a $\frac14$-gradient-stable approximation of $\|\cdot\|$ with error $(\alpha, \gamma)$ for $\e = \frac{1}{4 \widehat{\OPT}}$.
				
				\vspace{2pt}
        \For{each time $t$}
           \State Choose $x^{(t)} = \argmin_{x \in \{e_1,\ldots,e_k\}} \Psi\big(C^{(1)} x^{(1)} + \ldots + C^{(t-1)} x^{(t-1)} + C^{(t)} x\big)$.
        \EndFor
    \end{algorithmic}
    \end{algorithm}

	We prove that this algorithm is $O(\alpha + \gamma)$-competitive, which implies Theorem \ref{thm:loadBal}. For ease of notation, we define $y^{(t)}:= C^{(t)} x^{(t)}$ to be the cost vector incurred by the algorithm in step $t$, and $\Lambda^{(t)} := y^{(1)} + \ldots + y^{(t)}$ to be the cumulative cost vector. Moreover, let $y^{(t)}_*$ be the cost vector of the offline optimal solution at step $t$, and define $\Lambda^{(t)}_*  := y^{(1)}_* + \ldots + y^{(t)}_*$.

We first upper bound the total cost of the algorithm $\Psi(\Load^{(T)})$ with respect to the proxy function $\Psi$. Since we run a greedy algorithm, we have 
\[
\Psi(\Load^{(T)}) - \Psi(0) ~=~ \sum_{t=1}^T \Big(\Psi(\Load^{(t)}) - \Psi(\Load^{(t-1)})\Big) ~\leq~ \sum_{t=1}^T \Big(\Psi(\Load^{(t-1)}+y^{(t)}_*) - \Psi(\Load^{(t-1)})\Big).
\]
The smooth game inequality in \Cref{lemma:likesmoothgame} implies that for  $\zeta = \exp\left(\epsilon \lVert \Load_*^{(T)} \rVert +\delta\right)$, we have
\[
\sum_{t=1}^T \left(\Psi(\Load^{(t-1)} + y^{(t)}_*) -  \Psi(\Load^{(t-1)})\right) ~\leq~ \zeta \Psi(\Load_*^{(T)}) + \left(\zeta - 1\right) \left(\Psi(\Load^{(T)}) - \Psi(0) \right).
\]
So, in combination,
\[
\left(2 - \zeta \right) \left(\Psi(\Load^{(T)}) - \Psi(0) \right) ~\leq~ \zeta \Psi(\Load_*^{(T)}),
\]
or equivalently
\[
\Psi(\Load^{(T)}) ~\leq~ \frac{\zeta}{2 - \zeta} \Psi(\Load_*^{(T)}) + \Psi(0).
\]

Now using the approximation properties of $\Psi$, we can translate this into an upper bound of the true cost of the algorithm with respect to the original norm $\|\cdot\|$. That is, by definition of $\Psi$ we have $\Psi(\Load^{(T)}) \geq \| \Load^{(T)} \|$, $\Psi(\Load_*^{(T)}) \leq \alpha \| \Load_*^{(T)} \| + \frac{\gamma}{\e}$, and $\Psi(\Load^{(0)}) \leq \frac{\gamma}{\e}$. Therefore,
\[
\textrm{\alg's cost} ~=~ \| \Load^{(T)} \| ~\leq~ \frac{\zeta}{2 - \zeta} \bigg(\alpha \| \Load_*^{(T)} \| + \frac{\gamma}{\e}\bigg) + \frac{\gamma}{\e}.
\]

Since $\epsilon = \frac{1}{4 \widehat{\OPT}}$ and $\delta = \frac{1}{4}$, we have $\zeta = \exp\Big(\frac{\OPT}{4\widehat{\OPT}} + \frac{1}{4}\Big) \le \exp(\frac{1}{2}) < 1.7$. So,
	\begin{align*}
		\textrm{\alg's cost} ~=~ \| \Load^{(T)} \| ~\leq~ O\Big(\alpha \| \Load_*^{(T)} \| + \gamma\, \widehat{\OPT}\Big) ~\le~ O((\alpha + \gamma) \cdot \OPT).
	\end{align*}	
	This proves the approximation ratio of the algorithm, and implies \Cref{thm:loadBal}.
\end{proof}

Next, we see applications of our techniques to online algorithms with non-symmetric objectives.

\ifx \hasmain \undefined
  \end{document}
\fi


\ifx \hasmain \undefined
  \input{preamble}
  \begin{document}
\fi

\newcommand{\cI}{\mathcal{I}}

	\subsection{Online Vector Scheduling} \label{sec:vecSched}

	In this section we consider  \emph{Online Vector Scheduling} problem (\vecSched), which is related to \loadBal  but has a non-symmetric objective.
	In this problem we have $m$ machines with each machine having $r$ resources (coordinates). At time $t$, a job comes and reveals the vector load $\ell_{t,i} \in \R^r_+$ it incurs if assigned to machine $j$. At this point, the algorithm needs to select a machine $\sigma(t) \in [m]$ where this job is assigned. For each resource $k$, there is an ``inner'' norm $\|\cdot\|_{(k)}$ over $\R^m$ (used to aggregate the cost for this recourse over all machines), and the ``outer'' norm $\ell_\infty$ over $\R^r$ (used to aggregate the cost over all resources). Letting $\mathsf{load}(k) := (\sum_{t : \sigma(t) = i} \ell_{t,i})_i \in \R_+^m$ denote the vector load incurred on resource $k$ by the algorithm, the goal is to minimize
	\begin{align}
		\bigg\| \Big(\|\mathsf{load}(1)\|_{(1)}, ~\ldots~, \|\mathsf{load}(r)\|_{(r)}\Big) \bigg\|_{\infty}. \label{eq:vecSchedDef}
	\end{align}
	
	We show that the algorithm for \loadBal  can be used to obtain the following.
	
	\begin{theorem} \label{thm:vecSched}
		Consider the problem \vecSched with inner norms $\|\cdot\|_{(1)}, \ldots, \|\cdot\|_{(r)}$. If for every $k \in [r]$ the norm $\|\cdot\|_{(k)}$ has a $\frac{1}{4}$-gradient-stable approximation with error $(\alpha_k,\gamma_k)$, then there exists an algorithm for the problem with competitive ratio 
		\begin{align*}
			O\Big(\max_k \gamma_k ~+~ (\log r) \max_k \alpha_k \Big).
		\end{align*}
    \focs{Moreover, this algorithm is efficient given value and gradient oracle access to these  $\frac14$-gradient-stable approximations to norms.}
	\end{theorem}
	
	For the special case where the inner norms are all $\ell_{p_k}$'s this obtains an $O(\log m + \log r)$ competitive ratio, and for the more general case of monotone symmetric norms this gives an $O(\log^2 m + \log r)$-approximation for general monotone symmetric inner norms. In particular this proves \Cref{thm:vecSchedSym}.

	\begin{proof}[Proof of \Cref{thm:vecSched}]
		The idea is to embed \vecSched into an instance of \loadBal with a nested norm objective, and use the algorithm for the latter from \Cref{thm:loadBal}. More precisely, given an instance $\cI_{VS}$ of \vecSched we construct the following instance $\cI_{GLB}$ of \loadBal: $\cI_{GLB}$ has $m \cdot r$ machines, each indexed by $(i,k) \in [m] \times [r]$ (which corresponds to the $k$-th resource of machine $i$ in $\cI_{VS}$), and each job has $m$ processing options (corresponding to the assignment on one of the machines of $\cI_{VS}$). The processing option $i$ for job $t$ adds load $(\ell_{t,i})_k$ on machine $(i,k)$ for all $k \in [r]$, i.e., $C^{(t)}_{(i,k),i} = (\ell_{t,i})_k$ for all $k \in [r]$. Finally, this instance $\cI_{GLB}$ has as objective function the norm $\vertm{\cdot}$ defined as
		\begin{align*}
			\vertm{x} := \bigg\|\Big(\|A_1 x\|_{(1)},~\ldots~,\|A_r x\|_{(r)} \Big)  \bigg\|_{\infty},
		\end{align*}
		where $A_k$ is the matrix that projects a vector $x \in \R^{m \cdot r}$ (indexed by $(i,k) \in [m] \times [r]$) to the $m$-dimensional subvector vector $A_k x = (x_{(i,k)})_{i}$.
		
		By construction, scheduling job $t$ in machine $i$ in the original instance $\cI_{VS}$ corresponds precisely to choosing the processing option $i$ for job $t$ in the new instance $\cI_{GLB}$. Moreover, the corresponding schedules on the instances have the same cost with respect to the objective functions of their respective instances. 
		Thus, the online algorithm for approximating $\cI_{VS}$ is simply to construct the instance $\cI_{GLB}$ in an online fashion (which is easy to do), and to apply over it the algorithm from Theorem~\ref{thm:loadBal} to decide the processing options/assignments of the jobs.  
		
		For its competitive ratio, it suffices to show that  $\vertm{\cdot}$ admits a good gradient-stable approximation. Since $\vertm{\cdot}$ is a composition of norms, we use \Cref{lem:composition}: Recalling that the norm $\|\cdot\|_{\infty}$ over $r$ coordinates admits a $0$-gradient-stable approximation with error $(1,\ln r)$ (\Cref{lem:SM}) and using the assumption on the norms $\|\cdot\|_{(k)}$'s, we can employ \Cref{lem:composition} with $c=\max_k \alpha_k$ to directly get that $\vertm{\cdot}$ admits a $\frac{1}{4}$-gradient-stable approximation with error $(\alpha, \gamma)$, where 
		\begin{align*}
			\alpha = \max_k \alpha_k~~~~\textrm{and}~~~~\gamma = 2 (\max_k \alpha_k) \bigg[\frac{1}{\max_k \alpha_k}\,\|\gamma_1,\ldots,\gamma_r\|_{\infty} + \ln r \bigg] = 2 \max_k \gamma_k + 2 (\ln r) \max_k \alpha_k.
		\end{align*}

	Given the guarantee from \Cref{thm:loadBal}, our algorithm for \vecSched has competitive ratio $O(\alpha+\gamma) = O(\max_k \gamma_k + (\ln r) \max_k \alpha_k)$. 
	
	\focs{Finally, observe that \Cref{lem:composition} also implies that we have efficient access to the value and gradient oracles of the above $\frac{1}{4}$-gradient-stable approximation of $\vertm{\cdot}$. Thus, we can run the algorithm in Theorem~\ref{thm:loadBal} efficiently.}
	This concludes the proof of \Cref{thm:vecSched}. 
	\end{proof}
	
	

	\subsection{Online Generalized Assignment with Convex Costs}
	
	 Next we consider the \emph{Online Generalized Assignment with Convex Costs} problem (\onGAP) where the objective function is not directly a norm. In this problem there are $m$ machines and $T$ online jobs, each with $k$ processing options. However, each job $t$ has both a cost matrix $C^{(t)}$ and a duration matrix $D^{(j)}$, both $m \times k$, indicating the cost and duration vector incurred over the machines if we process the job with each of the $k$ options. There are two non-negative functions $f_{cost}, f_{dur}: \R^m \rightarrow \R_+$ that are convex, monotone, and symmetric. The goal is to choose the processing options for the jobs $x^{(1)},\ldots,x^{(T)} \in \{e_1,\ldots,e_k\}$ to minimize the sum of total cost and duration:
	\begin{align*}
		f_{cost}\Big(\sum_t C^{(t)} x^{(t)} \Big) + 		f_{dur}\Big(\sum_t D^{(t)} x^{(t)} \Big).
	\end{align*}
	Also, recall that a function $f$ has growth of order at most $p$ if $f(\alpha x) \le \alpha^p f(x)$ for all $\alpha \ge 1$ and for all vectors $x \in \R_+^m$.
	
	We show that we can get an $O(\log^{2p} m)$-competitive algorithm for this problem, where $p$ is an upper bound on the growth of $f_{cost}$ and $f_{dur}$.
	
    \onlineGAP*

	In the remainder of the section we describe the algorithm and then prove this result. Again, using the standard guess-and-double trick we can assume that  an estimate $\widehat{\OPT} \in [\OPT, 2\,\OPT]$ of the offline optimum $\OPT$ is given (this only adds an extra constant factor to the competitive ratio).	The high-level idea is to convert the objective function into a composition of norms associated with the sub-level sets of the functions $f_{cost}, f_{dur}$ of value at most $\widehat{\OPT}$, and then show that this composition admits a gradient-stable approximation so that we can apply the \loadBal algorithm from \Cref{thm:loadBal}.
	
	More precisely, given an instance $\cI_{onGAP}$, we construct the instance $\cI_{GLB}$ as follows: $\cI_{GLB}$ has $2m$ machines, each indexed by $(i,w) \in [m] \times [2]$, and each job has $k$ processing options. The processing option $j$ for job $t$ adds load $\mathsf{column}(C^{(t)}, j) \in \R_+^m$ to the machines $((i,1))_i$ and adds load $\mathsf{column}(D^{(t)}, j) \in \R_+^m$ to the machines $((i,2))_i$. Define the norms 
	\begin{align*}
		\|x\|_{cost} := \inf\Big\{ \lambda > 0 \mid f_{cost}\Big(\frac{x}{\lambda}\Big) \le \widehat{\OPT}\Big\} \quad\quad \textrm{and} \quad\quad \|x\|_{dur} := \inf\Big\{\lambda > 0 \mid f_{dur}\Big(\frac{x}{\lambda}\Big) \le \widehat{\OPT}\Big\}.
	\end{align*}
	Finally, the instance $\cI_{GLB}$ has  objective function the norm $\vertm{\cdot}$ defined as
		\begin{align*}
			\vertm{x} = \bigg\|\Big(\|A_1 x\|_{cost}~,~\|A_2 x\|_{dur} \Big)  \bigg\|_1,
		\end{align*}
		where as before $A_w$ is the matrix that projects a vector $x \in \R^{2m}$ (indexed by $(i,w) \in [m] \times [2]$) to the $m$-dimensional subvector vector $A_w x = (x_{(i,w)})_{i}$.
		
		The online algorithm for approximating $\cI_{onGAP}$ is simply to construct the instance $\cI_{GLB}$ above in an online fashion (which is easy to do), and to apply over it the algorithm from Theorem \ref{thm:loadBal} to decide the processing options of the jobs.

		We prove that this algorithm is $O(\log^{2p} m)$-competitive, as claimed in \Cref{thm:GAP}.
		
	\begin{proof}[Proof of \Cref{thm:GAP}]
		Let $x^{(1)}, \ldots, x^{(T)}$ and  $x^{(1)}_*,\ldots, x^{(T)}_*$ be the solution returned by the algorithm and the optimal offline solution for the instance $\cI_{onGAP}$ (with value $\OPT$), respectively. Let $\load^{(T)} = \sum_t C^{(t)} x^{(t)}$ and $M^{(T)} = \sum_t D^{(t)} x^{(t)}$, and define $\load^{(T)}_*$ and $M^{(T)}_*$ similarly with respect to the optimal solution. 
		
		Since $f_{cost}(\load^{(T)}_*) + f_{dur}(M^{(T)}_*) = \OPT \le \widehat{\OPT}$ and the functions are non-negative, we have that $\|\load^{(T)}_*\|_{cost}$ and $\|M^{(T)}_*\|_{dur}$ are at most 1. Therefore, we have  $$\bigg\|\Big(\|\load^{(T)}_*\|_{cost}~,~\|M^{(T)}_*\|_{dur} \Big)\bigg\|_1 \le 2,$$ and since the left-hand side is the value of the solution $x_*$ in the instance $\cI_{GLB}$, the optimum of the latter at most 2.
		
		Moreover, since the functions $f_{cost}$ and $f_{dur}$ are symmetric, the norms $\|\cdot\|_{cost}$ and $\|\cdot\|_{dur}$ are also symmetric. Therefore, from \Cref{thm:symm} we know that both admit a $\frac{1}{4}$-gradient-stable approximation with error $(O(\log m), O(\log^2 m))$. Also, from Lemma \ref{lemma:smoothLp} the $\ell_1$-norm admits a $0$-gradient-stable approximation with error $(1,0)$. Therefore employing \Cref{lem:composition} (with parameter $c = O(\log m)$) to the norm $\vertm{\cdot}$ we see that it admits a $\frac{1}{4}$-gradient-stable approximation with error $(\alpha,\gamma)$ where
		\begin{align*}
			\alpha = O(\log m)~~~~\textrm{and}~~~~\gamma = O(\log m) \bigg[\frac{1}{O(\log m)}\cdot 2 \cdot O(\log^2 m)\bigg] = O(\log^2 m).
		\end{align*}
		\focs{Moreover, since \Cref{thm:symm}  gives us efficient value and gradient oracles for $\frac14$-gradient-stable approximations of both $\|\cdot\|_{cost}$ and $\|\cdot\|_{dur}$, we also get from \Cref{lem:composition} efficient value and gradient oracles for $\frac{1}{4}$-gradient-stable approximation of $\vertm{\cdot}$.}
						
		 
		 Therefore, by the guarantee from Theorem \ref{thm:loadBal}, the solution $x^{(1)},\ldots,x^{(T)}$ is $O(\alpha + \gamma) = O(\log^2 m)$ competitive for the instance $\cI_{GLB}$, and so it has $\vertm{\cdot}$-value at most $2 \cdot O(\log^2 m) = O(\log^2 m)$, namely 
		 \[ 
		 \|\load^{(T)}\|_{cost} + \|M^{(T)}\|_{dur} ~=~ \bigg\|\Big(\|\load^{(T)}\|_{cost}~,~\|M^{(T)}\|_{dur} \Big)\bigg\|_1 ~\le~ O(\log^2 m).
		 \]
		 Unpacking the definitions, this implies in particular that $$f_{cost}\bigg(\frac{\load^{(T)}}{O(\log^2 m)} \bigg) ~\le~ \widehat{\OPT} ~\le~ 2 \OPT.$$ Then using the fact that the function $f_{cost}$ has growth of order at most $p$, this implies 
		 \begin{align*}
		 	f_{cost}(\load^{(T)}) = f_{cost}\bigg(O(\log^2 m) \cdot \frac{\load^{(T)}}{O(\log^2 m)} \bigg) \le O(\log^2 m)^p \cdot f_{cost}\bigg(\frac{\load^{(T)}}{O(\log^2 m)} \bigg) \le O(\log^{2p} m) \cdot \OPT. 
		 \end{align*}
		 For the same reasons we get $f_{dur}(M^{(T)}) \le O(\log^{2p} m) \cdot \OPT$, and thus the solution $x^{(1)},\ldots,x^{(T)}$ is $O(\log^{2p} m)$-competitive for the instance $\cI_{onGAP}$: $$f_{dur}(\load^{(T)}) + f_{dur}(M^{(T)}) ~\le~ O(\log^{2p} m) \cdot \OPT.$$ This concludes the proof of \Cref{thm:GAP}. 
	\end{proof}

\ifx \hasmain \undefined
  \end{document}
\fi

\section{Applications to Bandit Algorithms} \label{sec:banditAlgs}

In this section we use gradient stable approximations of norms to design bandit algorithms. 

\subsection{Approximate Converse to Jensen's Inequality} \label{sec:approxConvJensen}
Along with smoothness, a key technical observation to design our bandit algorithms is that the increase of $\Psi$ can be bounded in terms of a surrogate cost, which is  defined based on its gradient. Note that convexity of $\Psi$ immediately implies that $\langle \nabla \Psi (\Lambda^{(t-1)}), y^{(t)} \rangle \leq \Psi (\Lambda^{(t-1)} + y^{(t)}) - \Psi (\Lambda^{(t-1)})$. However, gradient stability  also implies the following reverse bound.  


\begin{lemma}
\label{lemma:PsiVsCost}
Consider any sequence of non-negative vectors $y^{(1)}, \ldots, y^{(T)} \in [0, 1]^d$. Furthermore, let $\Load^{(t)} = \sum_{s = 1}^t y^{(s)}$ be the sum of the first $t$ vectors. 
For any norm $\|\cdot\|$ that admits a $\delta$-gradient-stable approximation  with error $(\alpha, \gamma$), we have for every $\e > 0$ that
\[
\sum_{t=1}^T \langle \nabla \Psi_\e (\Lambda^{(t-1)}), y^{(t)} \rangle ~\geq~ \exp(-\epsilon \lVert \ones \rVert - \delta) \left( \Psi_\e(\Lambda^{(T)}) - \Psi_\e(\Lambda^{(0)}) - \Psi_\e(\ones) \right).
\]
\end{lemma}

\begin{proof}
Define $\tilde{\Lambda}^{(t)}_i = \max\{0, \Lambda^{(t)}_i - 1\}$ and $\tilde{y}^{(t)} = \tilde{\Lambda}^{(t)}_i - \tilde{\Lambda}^{(t-1)}_i$. Notice that $y^{(t)} \geq \tilde{y}^{(t)}$ component-wise for all $t$. This also implies that $\tilde{y}^{(t)}_i \leq 1$ for all $i$, so by monotonicity $\lVert \tau \tilde{y}^{(t)} \rVert \leq \lVert \ones \rVert$ for all $\tau \in [0, 1]$, which also implies $\nabla \Psi_\e (\tilde{\Lambda}^{(t-1)} + \tau \tilde{y}^{(t)}) \geq \exp(-\epsilon \lVert \ones \rVert - \delta) \nabla \Psi_\e (\tilde{\Lambda}^{(t-1)})$ by gradient stability. This implies
\begin{align*}
   \langle \Psi_\e (\Lambda^{(t-1)}), y^{(t)} \rangle ~ \geq~  \int_0^1 \langle \nabla \Psi_\e (\Lambda^{(t-1)}), \tilde{y}^{(t)} \rangle \,\mathrm{d}\tau ~&\geq~ \exp(-\epsilon \lVert \ones \rVert - \delta) \int_0^1 \langle \nabla \Psi_\e (\tilde{\Lambda}^{(t-1)} + \tau \tilde{y}^{(t)}), \tilde{y}^{(t)} \rangle \,\mathrm{d}\tau \\
   & =~ \exp(-\epsilon \lVert \ones \rVert - \delta) \left( \Psi_\e(\tilde{\Lambda}^{(t)}) - \Psi_\e(\tilde{\Lambda}^{(t-1)}) \right).
\end{align*}
By a telescoping sum, we get
$
\sum_{t=1}^T \langle \Psi_\e (\Lambda^{(t-1)}), y^{(t)} \rangle ~\geq~ \exp(-\epsilon \lVert \ones \rVert - \delta) \left( \Psi_\e(\tilde{\Lambda}^{(T)}) - \Psi_\e(\tilde{\Lambda}^{(0)}) \right).
$
Furthermore, by subadditivity  $\Psi_\e(\tilde{\Lambda}^{(T)}) \geq \Psi_\e(\Lambda^{(T)}) - \Psi_\e(\Lambda^{(T)} - \tilde{\Lambda}^{(T)}) \geq \Psi_\e(\Lambda^{(T)}) - \Psi_\e(\ones)$ and by definition $\Psi_\e(\tilde{\Lambda}^{(0)}) = \Psi_\e(0) = \Psi_\e(\Lambda^{(0)})$. This implies the lemma.
\end{proof}

We now have the tools for our bandit applications.

\subsection{Bandits with Knapsacks}
We will first consider the following \emph{Bandits with Knapsacks} (\BwK) problem for adversarial arrivals. Here, an algorithm chooses one of $n$ actions in every time step. Each action gives a reward and incurs a vector load. The process stops when a norm of the sum of vector loads exceeds some given budget $B$, or when the time horizon $T$ is reached.

In more detail, an adversary initially chooses $T$ reward vectors $r^{(1)}, \ldots, r^{(T)} \in [0,1]^d$ and $T$ load matrices $C^{(1)}, \ldots, C^{(T)} \in [0,1]^{d \times n}$, which are unknown to the algorithm. In time step $t$, the algorithm chooses an action, where the $i$-th action gives reward $r^{(t)}_i \in [0,1]$ and incurs a vector load of $C^{(t)} e_i \in [0,1]^d$, both of which are unknown before playing the action. Importantly, there is also a null action, not causing any reward nor load. We let $x^{(t)} \in \{e_1, \ldots, e_n\}$ denote the vector indicating which action is chosen in step $t$. As soon as the algorithm reaches a point such that $\lVert \sum_{t' < t} C^{(t')} x^{(t')} \rVert > B$, only the null action can be chosen. After the algorithm has chosen the action, it only gets to know $\langle r^{(t)}, x^{(t)} \rangle$ and $C^{(t)} \cdot x^{(t)}$ (bandit feedback).

The benchmark for \BwK is defined as follows: For any fractional choice $x^\ast \in \Delta_n := \{ x \in [0, 1]^n \mid \sum_{i=1}^n x_i = 1 \}$ of the $n$ actions, we let $\tau^\ast = \min\{ t \mid \lVert \sum_{t'=1}^t C^{(t)} \cdot x^\ast \rVert > B\}$ be the time step at which $x^\ast$ would run out of budget, or $\tau^\ast =T$ if there is budget left. Then the optimum $\OPTBwK$ is defined the maximum over all $\sum_{t=1}^{\tau^\ast} \langle r^{(t)}, x^\ast \rangle$.

\thmbwK*

The assumption that $\OPTBwK$ is known can be removed in this theorem at a further multiplicative loss of $O(\log T)$, which is known to be unavoidable~\cite{ISSS-FOCS19}.

		\begin{algorithm}[H]
		\caption{Bandits with Knapsacks}
		\begin{algorithmic}[0]      
				\vspace{-2pt}
				\State \textbf{input:} Estimate $\OPTBwK$
				\vspace{2pt}
				\State Let $\Psi=\Psi_\e$ be a $\frac14$-gradient-stable approximation of $\|\cdot\|$ with error $(\alpha, \gamma)$ at scale $\e = \frac{4(\alpha+\gamma)}{B}$.
				
				\vspace{2pt}
        \For{each time $t$}
           \State Adopt choice of action $x^{(t)}$ from bandits algorithm (e.g. \textsc{Exp3.P}).
           \State Return $\langle \mathcal{R}^{(t)}, x^{(t)} \rangle$ to bandits algorithm as the reward as defined by Equations~\eqref{eq:lagrangianreward} and \eqref{eq:lambda}.
        \EndFor
    \end{algorithmic}
    \label{alg:bwk}
    \end{algorithm}

Our algorithm, \Cref{alg:bwk}, generalizes the one in \cite{KS-COLT20}. It uses a $\frac14$-gradient-stable approximation $\Psi=\Psi_\e$ of $\|\cdot\|$ with error $(\alpha, \gamma)$ for $\e = \frac{4(\alpha+\gamma)}{B}$. Note that this choice of $\e$ is feasible because $\alpha$ and $\gamma$ do not depend on $\e$. Based on this norm approximation, it defines a surrogate game, which can be viewed as a Lagrangian relaxation of the original problem. In this surrogate game it applies a classic no-regret bandits algorithm, which is allowed to choose actions freely without any constraints, such as \textsc{Exp3.P} \cite{DBLP:journals/siamcomp/AuerCFS02}. Generally, we do not need any property besides a bound on the regret with probability $1-p$ against an adaptive adversary for an unknown time horizon.

Specifically, the $i$-th action's surrogate reward in time step $t$ is defined as
\begin{equation}
\label{eq:lagrangianreward}
\mathcal{R}^{(t)}_i = \begin{cases} r^{(t)} e_i - \lambda 	\cdot \langle \nabla \Psi (\load^{(t-1)}), C^{(t)} \cdot e_i \rangle & \text{ if $\lVert \load^{(t-1)} \rVert \leq B$} \\
0 & \text{ otherwise,}
\end{cases}
\end{equation}
where $r^{(t)}$ is the reward vector from BwK problem,  $\load^{(t-1)} = \sum_{s = 1}^{t-1} C^{(s)} \cdot x^{(s)}$ is the load after time~$t-1$, and
\begin{equation}
\label{eq:lambda}
\lambda :=  \frac{\sum_{t = 1}^{\tau^\ast} \langle r^{(t)}, x^\ast \rangle}{6 e^5 (\alpha + \gamma) B} =  \frac{\OPTBwK}{6 e^5 (\alpha + \gamma) B}.
\end{equation}
Note that in order to define $\lambda$, our algorithm has to know the value of $\OPTBwK$, which we assumed to be part of the input. 

In order to bound the regret an algorithm could guarantee in this surrogate game, observe that by Lemma~\ref{lemma:gradientBound}, we have $\langle C^{(t)} e_i, \nabla \Psi (\load^{(t-1)}) \rangle \leq  \alpha \| \one \|$ and consequently $\lvert \mathcal{R}^{(t)}_i \rvert \leq \rho$ for $\rho := 1+\lambda\cdot \alpha \cdot \| \one \|$. Therefore, we can we can guarantee that
\begin{equation}
\label{eq:bwksurrogateregret}
\sum_{t=1}^{\tau^\ast} \langle \mathcal{R}^{(t)}, x^{(t)} \rangle ~\leq~ \sum_{t=1}^{\tau^\ast} \langle \mathcal{R}^{(t)}, x^\ast \rangle + \rho \cdot \textsc{Regret}_{\tau^\ast},
\end{equation}
where $\textsc{Regret}_{\tau^\ast}$ would be the regret of the algorithm against an adaptive adversary choosing losses in $[-1, 1]$. Note that $\tau^\ast$ is not known to the algorithm. However, it is a fixed number not depending on the choices of the algorithm. So it is sufficient to apply any algorithm operating with an unknown time horizon. Here, \textsc{Exp3.P} \cite{DBLP:journals/siamcomp/AuerCFS02} would give us $\textsc{Regret}_{\tau^\ast} = O(\sqrt{\tau^\ast n \log (n/p)})$ with probability $1-p$.

\focs{Observe that the algorithm only needs one gradient query of $\Psi(\cdot)$ per time step.}
Now, it remains to show that the algorithm obtains the claimed reward.

\begin{proposition} \label{prop:BwKAdversarial}
Whenever Equation~\eqref{eq:bwksurrogateregret} is fulfilled and $B \geq 4 \cdot (\alpha + \gamma) \cdot \| \ones \|$, the reward obtained by actions $x^{(1)}, \ldots, x^{(T)}$ in the BwK problem fulfills
\[
\sum_{t = 1}^\tau \langle r^{(t)}, x^{(t)} \rangle ~~\geq~~ \Omega\Big(\frac{1}{\alpha + \gamma}\OPTBwK\Big) - O\Big(\frac{\OPTBwK \cdot \| \one \| }{(\alpha + \gamma) \cdot B}\Big)\cdot \textsc{Regret}_{\tau^\ast}.
\]
\end{proposition}

\begin{proof}
We distinguish the analysis in two cases: Either the algorithm stops before  time $\tau^\ast$ (recall, this is time at which  $x^\ast$ runs out of budget, otherwise $\tau^\ast = T$) or it stays within budget until $\tau^\ast$.

\paragraph{Case 1: $\Psi(\load^{(\tau^\ast)}) > B$.}

As $\Psi(\load^{(\tau^\ast)}) > B$ implies that $\lVert \load^{(\tau^\ast)} \rVert > B$, there is some time $\tau \leq \tau^\ast$ at which $\lVert \load^{(\tau)} \rVert > B$ for the first time. This means that $\tau$ is the last round before the algorithm stops and $\mathcal{R}^{(t)} = 0$ for all $t > \tau$. We compare to always playing the null action, which has surrogate reward $0$. By Equation~\eqref{eq:bwksurrogateregret}, in the surrogate game, we therefore get
\[
\sum_{t = 1}^{\tau} \langle \mathcal{R}^{(t)}, x^{(t)} \rangle ~=~ \sum_{t = 1}^{\tau^\ast} \langle \mathcal{R}^{(t)}, x^{(t)} \rangle ~\geq~ 0 - \rho \cdot \Regret_{\tau^\ast}.
\]
Using the definition of $\mathcal{R}^{(t)}$, this implies
\begin{align*}
\sum_{t = 1}^{\tau} \langle r^{(t)}, x^{(t)} \rangle ~\geq~ \lambda \sum_{t = 1}^{\tau} \langle \nabla \Psi(\load^{(t-1)}), C^{(t)} x^{(t)} \rangle - \rho \cdot \Regret_{\tau^\ast}.
\end{align*}
Since $\Psi$ approximates the norm, we have $\Psi(\load^{(\tau)}) - \Psi(0) - \Psi(\ones) \geq B - (\alpha \lVert \ones \rVert + \frac{\gamma}{\epsilon}) - (\alpha \cdot 0 + \frac{\gamma}{\epsilon}) \geq \frac{B}{4}$, where in the last step we use the definition of $\epsilon$ and that $B \geq 4 (\alpha + \gamma) \lVert \ones \rVert$.

So, by Lemma~\ref{lemma:PsiVsCost},
\begin{align*}
\sum_{t = 1}^{\tau} \langle \nabla \Psi(\load^{(t-1)}), C^{(t)} x^{(t)} \rangle ~& \geq~ \exp(-\epsilon \lVert \ones \rVert -\delta) (\Psi(\load^{(\tau)}) - \Psi(0) - \Psi(\ones))\\
& \geq~ \frac{B}{4}\exp(-\epsilon \lVert \ones \rVert -\delta) ~ \geq~ \frac{B}{4} \exp(-5).
\end{align*}
This implies
\begin{align*} 
\sum_{t = 1}^{\tau} \langle r^{(t)}, x^{(t)} \rangle ~\geq~ \lambda \cdot \frac{B}{4 e^5} - \rho \cdot \Regret_{\tau^\ast} ~\geq~ \frac{\OPTBwK}{24 e^{10} (\alpha + \gamma)} - \rho \cdot \Regret_{\tau^\ast}.
\end{align*}

\paragraph{Case 2: $\Psi(\load^{(\tau^\ast)}) \leq B$.}

In this case, $\lVert \load^{(t)} \rVert \leq \Psi(\load^{(\tau^t)}) \leq B$ for all $t \leq \tau^\ast$. As an auxiliary point of comparison, we now use a scaled-down version of the optimal choice, namely, $x' := \frac{1}{\alpha + \gamma}  x^\ast$. Note that also $x'$ is a feasible fractional solution because we assume that there is a null action. By Equation~\eqref{eq:bwksurrogateregret}, we get
$\sum_{t = 1}^{\tau^\ast} \langle \mathcal{R}^{(t)}, x^{(t)} \rangle \geq \sum_{t = 1}^{\tau^\ast} \langle \mathcal{R}^{(t)}, x' \rangle - \rho \cdot \Regret_{\tau^\ast}$, and therefore by the definition of $\mathcal{R}^{(t)}$,
\begin{align*}
\sum_{t = 1}^{\tau^\ast} \langle r^{(t)}, x^{(t)} \rangle & \geq \sum_{t = 1}^{\tau^\ast} \left( \langle r^{(t)}, x' \rangle - \lambda \cdot \langle \nabla \Psi(\load^{(t-1)}), C^{(t)} x' \rangle + \lambda \cdot \langle \nabla \Psi(\load^{(t-1)}), C^{(t)} x^{(t)} \rangle \right)  - \rho \cdot \Regret_{\tau^\ast} \\
& \geq \frac{\OPTBwK}{\alpha + \gamma} - \lambda \Big( \sum_{t = 1}^{\tau^\ast} \langle \nabla \Psi(\load^{(t-1)}), C^{(t)} x' \rangle \Big) - \rho \cdot \Regret_{\tau^\ast},
\end{align*}
where the last step uses that $\sum_{t = 1}^{\tau^\ast} \langle r^{(t)}, x' \rangle = \frac{\OPTBwK}{\alpha + \gamma}$ and that all entries in $C^{(t)}$ as well as the gradients of $\Psi$ are non-negative.

Define $y^{(t)}_* = \frac{1}{\alpha + \gamma} C^{(t)} x'$ and $y^{(t)} = C^{(t)} x^{(t)}$. By convexity of $\Psi$, we have
\[
\sum_{t = 1}^{\tau^\ast} \langle \nabla \Psi(\load^{(t-1)}), y^{(t)}_* \rangle ~\leq~ \sum_{t=1}^{\tau^\ast} \left( \Psi(\Load^{(t-1)} + y^{(t)}_*) - \Psi(\Load^{(t-1)}) \right).
\]

Now, we use Lemma~\ref{lemma:likesmoothgame}, replacing $T$ by $\tau^\ast$ in the statement, and the fact that $\Psi(\Load_*^{(\tau^\ast)}) \leq \alpha \frac{B}{\alpha + \gamma} + \frac{\gamma}{\epsilon} \leq 2 B$ and that $\Psi(\Load^{(\tau^\ast)}) \leq B$. This gives us that
\[
\sum_{t=1}^{\tau^\ast} \left( \Psi(\Load^{(t-1)} + y^{(t)}_*) - \Psi(\Load^{(t-1)}) \right) ~\leq~ \exp\left(\epsilon \frac{B}{\alpha + \gamma} +\delta\right) 3 B ~\leq~ 3 e^5 B.
\]
So, we get
\[
\sum_{t = 1}^{\tau^\ast} \langle r^{(t)}, x^{(t)} \rangle ~\geq~ \frac{\OPTBwK}{\alpha + \gamma} - \lambda \cdot 3 e^5 B - \rho \cdot \Regret_{\tau^\ast} ~=~ \frac{\OPTBwK}{6 e^5 (\alpha + \gamma)} - \rho \cdot \Regret_{\tau^\ast}. \qedhere
\]
\end{proof}

\begin{remark}
If the algorithm gets to know the entire vector $r^{(t)}$ and the entire matrix $C^{(t)}$ after step $t$ the bound in \Cref{theorem:bwk} can be improved to $\textsc{Regret} = O(\sqrt{T \log n})$, namely by using \textsc{Hedge} \cite{FS-GEB99}.
\end{remark}

%


\ifx \hasmain \undefined
  \input{preamble}
  \begin{document}
\fi


\subsection{Bandits with Vector Costs} \label{sec:BwVC}


In this section we consider the Bandits with Vector Costs (\BwVC) problem, which is a natural generalization of the classical bandits problem when the actions incur vector costs. Here, in each time step $t$, an algorithm can choose one of $n$ actions (``experts''). In each time step, each action has a $d$-dimensional cost vector associated to it and the goal is to minimize a norm on the sum of cost vectors. That is, an adversary defines $T$ matrices $C^{(1)}, \ldots, C^{(T)} \in [0, 1]^{d \times n}$ and the algorithm chooses vectors $x^{(1)}, \ldots, x^{(T)} \in \{0,1\}^n$, each containing exactly one entry that is $1$, attempting to minimize $\lVert \sum_{t=1}^T C^{(t)} \cdot x^{(t)} \rVert$. The key difference to generalized load balancing is that the algorithm only gets to know $\langle C^{(t)}, x^{(t)} \rangle$ only after it has chosen $x^{(t)}$.

The benchmark is the best fractional choice of actions. That is, we consider $x^\ast \in \Delta_n := \{ x \in [0, 1]^n \mid \sum_{i=1}^n x_i = 1 \}$ so that $\rVert \sum_{t=1}^T C^{(t)} \cdot x^\ast \rVert$ is minimized.

We define $\Load^{(t)} = \sum_{s=1}^t C^{(t)} \cdot x^{(t)}$ as sum of the vector costs incurs in the first $t$ steps and $\Load^{(t)}_* = \sum_{s=1}^t C^{(t)} \cdot x^\ast$ the respective quantity for the benchmark solution.

\newcounter{foo1}
\setcounter{foo1}{\value{section}}
\newcounter{foo2}
\setcounter{foo2}{\value{theorem}}

\setcounter{section}{1}
\setcounter{theorem}{\value{banditsVC}}

\begin{theorem} \label{thm:bwVCGeneral}
	Consider the problem Bandits with Vector Costs with a norm $\|\cdot\|$. If $\|\cdot\|$ admits a $\frac{1}{4}$-gradient-stable approximations with error $(\alpha,\gamma)$, then there exists an algorithm that guarantees $\lVert \Load^{(T)} \rVert = O((\alpha + \gamma) \cdot \lVert \Load^{(T)}_* \rVert + \alpha \cdot \lVert \ones \rVert \cdot \textsc{Regret})$ with probability $1-p$, where $\textsc{Regret} = O(\sqrt{T n \log(n/p)})$ and $p \in [0,1]$ is a parameter. \focs{Moreover, this algorithm is efficient given gradient oracle access to this $\frac{1}{4}$-gradient-stable approximation.}
 \end{theorem}
 
 \setcounter{section}{\value{foo1}}
 \setcounter{theorem}{\value{foo2}}
 

Our algorithm for \BwVC reduces the problem to classic online learning with a scalar cost. It then adopts the action chosen by this algorithm. To this end it uses a gradient-stable approximation of the norm as follows. In each step, we define a surrogate linear cost function by defining $c^{(t)} = (\nabla \Psi (\Lambda^{(t-1)}))^\mathrm{transposed} \cdot C^{(t)}$. That is, the surrogate scalar cost of $x$ is $\langle c^{(t)}, x \rangle = \langle \nabla \Psi (\Lambda^{(t-1)}), C^{(t)} \cdot x^{(t)} \rangle$. So, we approximate $\Psi$ by its gradient at the current load $\Lambda^{(t-1)}$.

By Lemma~\ref{lemma:gradientBound} and because $C^{(t)} \in [0,1]^{d \times n}$, we always have $c^{(t)} \in [0, \rho]$ for $\rho := \alpha \lVert \ones \rVert$. Therefore, applying any no-regret learning algorithm for a scalar cost problem. Using \textsc{Exp3.P} \cite{DBLP:journals/siamcomp/AuerCFS02} we are guaranteed that with probability $1-p$
\begin{equation}
\label{eq:olvc-surrogate-regret}
\sum_{t = 1} \langle c^{(t)}, x^{(t)} \rangle \leq \sum_{t = 1} \langle c^{(t)}, x^\ast \rangle + \rho \cdot \textsc{Regret},
\end{equation}
where $\textsc{Regret} = O(\sqrt{T n \log(n/p)})$.

\focs{Observe that the algorithm only needs one gradient query of $\Psi$ per time step.}
Now it only remains to prove the following proposition in order to show the theorem.

\begin{proposition}
Whenever \eqref{eq:olvc-surrogate-regret} is satisfied, $\| \Load^{(T)} \| = O( (\alpha + \gamma) \cdot \| \Load_*^{(T)} \| + \rho \cdot \textsc{Regret})$.
\end{proposition}

\begin{proof}
Letting $y^{(t)}_* = C^{(t)} \cdot x^*$, we have by convexity of $\Psi$
\[
\langle \nabla\Psi(\Lambda^{(t-1)}), y^{(t)}_* \rangle \leq \Psi(\Lambda^{(t-1)} + y^{(t)}_*) - \Psi(\Lambda^{(t-1)});
\]
so also by the definition of $c^{(t)}$ and by Lemma~\ref{lemma:likesmoothgame}
\[
\sum_{t=1}^T \langle c^{(t)}, x^* \rangle ~\leq~ \sum_{t=1}^T \Psi(\Lambda^{(t-1)} + y^{(t)}_*) - \Psi(\Lambda^{(t-1)}) ~\leq~ \zeta \Psi(\Load_*^{(T)}) + \left(\zeta - 1\right) \left(\Psi(\Load^{(T)}) - \Psi(\Load^{(0)}) \right),
\]
where $\zeta = \exp\left(\epsilon \lVert \Load_*^{(T)} \rVert +\delta\right)$.  

Combining this bound with Lemma~\ref{lemma:PsiVsCost} and the regret property, we get
\begin{align*}
\Psi(\Lambda^{(T)}) - \Psi(\Lambda^{(0)}) - \Psi(\ones) & \leq \exp(\epsilon \lVert \ones \rVert + \delta) {\textstyle \sum_{t=1}^T} \langle c^{(t)}, x^{(t)} \rangle \\
&\leq \exp(\epsilon \lVert \ones \rVert + \delta) \Big( \rho \cdot \textsc{Regret} + {\textstyle \sum_{t=1}^T} \langle c^{(t)}, x^* \rangle \Big) \\
& \leq \exp(\epsilon \lVert \ones \rVert + \delta) \left( \rho \cdot \textsc{Regret} +  \zeta \Psi(\Load_*^{(T)}) + \left(\zeta - 1\right) \left(\Psi(\Load^{(T)}) - \Psi(\Load^{(0)}) \right) \right).
\end{align*}

This implies
\begin{align*}
& (2 - \exp(\epsilon \lVert \ones \rVert + \epsilon \lVert \Load_*^{(T)} \rVert + 2 \delta)) \left(\Psi(\Load^{(T)}) - \Psi(\Load^{(0)}) \right) \\
& \qquad \qquad \leq \quad \exp(\epsilon \lVert \ones \rVert + \epsilon \lVert \Load_*^{(T)} \rVert + 2 \delta)\left( \Psi(\Load_*^{(T)}) + \rho \cdot \textsc{Regret}\right) + \Psi(\ones).
\end{align*}

For $\epsilon := 1/(4\cdot \lVert \Load^{(T)}_*  \rVert + 4 \lVert \ones  \rVert)$, $\delta := 1/8$, we have $\exp(\epsilon\lVert \Load^{(T)}_* \rVert+\epsilon \lVert \ones \rVert+2\delta)
\leq \exp(1/4 + 1/4) \leq \frac{9}{5}$. And thus,
\[
\frac{1}{5} \left(\Psi(\Load^{(T)}) - \Psi(\Load^{(0)}) \right) ~\leq~ \frac{9}{5} \left( \Psi(\Load_*^{(T)}) + \rho \cdot \textsc{Regret}\right) + \Psi(\ones),
\]
or equivalently,
$
\Psi(\Load^{(T)}) ~\leq~ 9 \cdot \Psi(\Load_*^{(T)}) + \Psi(\Load^{(0)}) + 5 \cdot \Psi(\ones) + 9 \cdot \rho \cdot \textsc{Regret}.
$

By the approximation properties of $\Psi$, we get
\[
\| \Load^{(T)} \| ~\leq~ 9 \cdot \alpha \cdot \| \Load_*^{(T)} \| + 5 \cdot \alpha \cdot \lVert \ones \rVert + 15 \cdot \frac{\gamma}{\epsilon} + 9 \cdot \rho \cdot \textsc{Regret}.     \qedhere
\]
\end{proof}

\begin{remark}
If the algorithm gets to know the entire matrix $C^{(t)}$ after step $t$, the bound in \Cref{thm:bwVCGeneral} can be improved to $\textsc{Regret} = O(\sqrt{T \log n})$ by using \textsc{Hedge} \cite{FS-GEB99}.
\end{remark}

\ifx \hasmain \undefined
  \end{document}
\fi

\section{Conclusions and Further Directions}

Our gradient-stable norm approximation gives 
$O(\log^2 d)$-competitive algorithms for several online and bandit problems, in particular for Online Generalized Load Balancing and Bandits with Knapsacks for symmetric norms. There are still many open problems and future directions. For example, we only know $\Omega(\log d)$  lower bounds for both these problems, which leaves open what are the tight competitive ratios. Another interesting direction is to consider other online problems with (symmetric) norm objectives, e.g. Online Set Cover \cite{cvxPDFOCS,vishSetCoverLp}, and obtain $\poly\!\log d$-competitive algorithms. 
Besides our approach via gradient-stable norm approximation, it would be also interesting to extend other techniques such as   online primal-dual  (see~\cite{BuchbinderNaor-Book09}) to online problems with symmetric norms.
Finally, it is an intriguing question to better understand arbitrary monotone norms and whether they admit good gradient-stable approximations.


\medskip
\noindent
\subsection*{Acknowledgments}
We are grateful to the anonymous reviewers of SODA 2023 for their helpful comments.

\bigskip 
\bigskip 

\appendix
\section{Further Related Work and Missing Proofs}
\ifx \hasmain \undefined
  \input{preamble}
  \begin{document}
\fi

\subsection{Further Related Work} \label{sec:further}

	\paragraph{Other notions of ``smoothness''.} ``Smoothness'' of functions (in a broad sense) has been recognized to play a very important role on an multitude of settings, and a comprehensive survey about the subject is impossible. Nonetheless, we briefly discuss notions that are most closely related to the gradient-stability  property
\begin{align}
	\nabla \Psi(x+y) \ge e^{-\e \|y\| - \delta} \cdot \nabla \Psi(x),~~~~~~~~\forall x,y \in \R^d_+.  \label{eq:gradStabApp}
\end{align}
The most classic such notion is that of a function $\Psi$ having Lipschitz gradient, namely $$\|\nabla \Psi(x+y) - \nabla \Psi(x)\|_{(1)} \le L \cdot \|y\|_{(2)},~~~~~~~~\forall x,y$$ for some norms $\|\cdot\|_{(1)}, \|\cdot\|_{(2)}$ and a constant $L$. A main difference between this and gradient-stability \eqref{eq:gradStabApp} is that in the latter we have a \textbf{multiplicative} bound between $\nabla \Psi(x+y)$ and $\nabla \Psi(x)$; this seems to be critical for the ``error'' between these gradients not to accumulate as we add over all the time steps in the analysis of online problems. 

	Another classic notion from the theory of Banach spaces is the so-called \emph{uniform smoothness} of norms~\cite{lindTza}. However, if $\|\cdot\|$ is uniformly smooth (and twice differentiable, for simplicity) we only seem to obtain a bound of the form $y^\top (\nabla^2 \|x\|) y \le C \|y\|^2$ for some constant $C$ and all $x,y$, which again only means that $\|\cdot\|$ has Lipschitz gradients. 
	
	Notions of ``smoothness'' of the so-called \emph{baseline potential} (which translate to stability of decisions) have also played a fundamental role in determining the regret in online learning algorithms~\cite{abernethySmoothing,abernethyNewSmoothness,simultaneousLoadBal,KS-COLT20,curvedOpt}. Indeed, in the standard algorithm FTRL algorithm the addition of the strongly convex regularizer translates to making the baseline potential strongly smooth, or equivalently, having Lipschitz gradients. In addition,~\cite{simultaneousLoadBal} also explicitly used the multiplicative smoothing condition 
	\begin{align}
	\nabla \Psi(x+y) \in e^{\pm \e \|y\|_{\infty}} \cdot \nabla\Psi(x),~~~~~~\forall x,y \in \R^d_+,   \label{eq:oldGradStab}
	\end{align}
	 which is a special case of gradient-stability where the length of the change vector $y$ is \textbf{always} measured in $\ell_\infty$-norm (this is implicitly present in the original FTPL analysis~\cite{KalaiVemp-Journ05}; see also~\cite{abernethyNewSmoothness} for a related notion). As discussed in Section \ref{sec:introGSApprox}, such approximation to a norm $\|\cdot\|$ can be obtained by setting $\Psi(x) = \E_\nu \|x + \nu\|$ where each coordinate of $\nu$ is an independent Exponential random variable with mean $1/\e$. However, also as discussed there, all known approximations satisfying \eqref{eq:oldGradStab} have a large error $\Psi(\zero) \gg \|0\|$ and yield online algorithms with very poor approximation guarantees. This weakness can be traced to the fact the guarantee in \eqref{eq:oldGradStab} is not tailored to the specific norm $\|\cdot\|$. 
	 
	 Finally,~\cite{oneSidedSmooth} introduced the closely related notion of \emph{one-sided smoothness} in the context of DR-submodular optimization. Adapted to convex functions, this definition requires a (twice differentiable) function to satisfy $y^\top \nabla^2 \Psi(x) y \ge - C \frac{\|y\|_1}{\|x\|_1} \ip{y}{\nabla \Psi(x)}$, noting that the bound is based specifically on the $\ell_1$-norm. In contrast, gradient-stability (with $\delta = 0$) implies $y^\top \nabla^2 \Psi(x) y \ge - \e \|y\| \ip{y}{\nabla \Psi(x)}$, with the bound depending on the norm $\|\cdot\|$; as in the previous paragraph, this tailoring to the given norm $\|\cdot\|$ seems to be crucial for obtaining low-error approximations. In addition,~\cite{oneSidedSmooth} does not discuss how to approximate a given function/norm by a one-sided smooth function.

	\paragraph{Other related results in online algorithms.} There are other results on online algorithm for problems with convex objectives that are related to our present work. However most of them assume that the objective function $f$ has monotone gradients, namely $\nabla f(x') \ge \nabla f(x)$ for all $x' \ge x \ge 0$~\cite{BGMS11,huangKim,cvxPDFOCS,fazelConic,onlineConvexPD,sodaMixed}. Notice that monotone norms are far from satisfying monotone gradients property, e.g., $\nabla \|(1,0)\|_2 = (1,0)$ but $\nabla \|(1,1)\|_2 = (\frac{1}{\sqrt{2}}, \frac{1}{\sqrt{2}})$. The only exception to  gradient monotonicity  that we know is~\cite{vishSetCoverLp}, where the authors consider a version of the Online Set Cover problem where the objective function is a sum of $\ell_p$ norms. However, in this case we have the crucial property that the function $\|\cdot\|_p^p$ does have monotone gradients, a fact that is crucially exploited in their analysis. 
	
	Also notice that gradient-stability can be thought of as an appropriately defined relaxation of gradient monotonicity that is still enough to yield algorithms with good approximation guarantees. 

\ifx \hasmain \undefined
  \end{document}
\fi

\subsection{Softmax and \texorpdfstring{$\ell_p$}{l\_p} Norms} \label{sec:lpNorms}

	Recall that $\nabla_i f$ denotes the $i$-th partial derivative of a function $f$. 

	\paragraph{Softmax function.} A standard ``smooth'' approximation of the $\ell_\infty$-norm is the \emph{softmax} function $SM_\e(x) := \frac{1}{\e} \ln \Big(\sum_i e^{\e x_i}\Big).$ The next lemma shows that the softmax actually provides a gradient-stable approximation.

	\begin{lemma}	\label{lem:SM}
		The softmax function provides a $0$-gradient-stable approximation of $\|\cdot\|_{\infty}$ (in $d$-dimensions) with error $(1, \ln d)$.
	\end{lemma}
	
	\begin{proof}
	Fix $\e > 0$ throughout. First, it is clear that the function $SM_\e$ is monotone, and it is well-known that it is also convex. It is also subadditive:
	\begin{align*}
		SM_\e(x+y) = \frac{1}{\e} \ln \bigg(\sum_i e^{\e x_i} e^{\e y_i}\bigg) \le \frac{1}{\e} \ln \bigg(\Big(\sum_i e^{\e x_i}\Big) \cdot \Big(\sum_i e^{\e y_i}\Big)\bigg) = SM_\e(x) + SM_\e(y).
	\end{align*}
	
	 For the approximation error, we have $SM_\e(x) = \frac{1}{\e} \ln \big(\sum_i e^{\e x_i}\big) \geq \frac{1}{\e} \ln \big(e^{\e \|x\|_\infty}\big) =  \|x\|_\infty$. In the other direction, we have $SM_\e(x) = \frac{1}{\e} \ln \big(\sum_i e^{\e x_i}\big) \leq \frac{1}{\e} \ln \big(d \cdot e^{\e \|x\|_\infty }\big) = \|x\|_\infty + \frac{\ln d}{\e}$. 
	
	Next, we compute the gradients: for every $i \in [d]$
		\[ \nabla_i SM_\e(x) ~=~  \frac{ e^{\e x_i}}{\sum_j e^{\e x_j}} \quad \text{ and } \quad 
		\nabla_i SM_\e(x+y) =    \frac{ e^{\e x_i + \e y_i}}{\sum_j e^{\e x_j+ \e y_j}} \geq \frac{ e^{\e x_i}}{(\sum_j e^{\e x_j})  \cdot  e^{\e \|y\|_\infty}}.
	\]
	Thus, we get
	\[ 	\nabla_i SM_\e(x+y) ~\geq~ \exp(-\e \|y\|_\infty) \cdot \frac{ e^{\e x_i}}{\sum_j e^{\e x_j}} ~=~ \exp(-\e \|y\|_\infty) \cdot \nabla_i SM_\e(x) .
	\]
	
	Since these bounds hold for every $\e > 0$, it proves the desired result. 
	\end{proof}


\paragraph{$\ell_p$ Norms.} The more general $\ell_p$-norms with $p \in [1,\infty)$ also admit gradient-stable approximations. 
 
\begin{lemma}	\label{lemma:smoothLp}
	For every $p \in [1,\infty)$, the $\ell_p$-norm admits a $0$-gradient-stable approximation with error $(1,\min\{p-1, \ln d\})$.
\end{lemma}

\begin{proof}
	Fix $\e > 0$ and define the approximation $\Psi(x) = \|x+c\ones\|_p - c$, where $c=\frac{p-1}{\epsilon \|\ones\|_p}$. 
	
	We see that $\Psi$ is monotone, convex (since the $\ell_p$-norm is convex); it is also subadditive, since by triangle inequality 
	\begin{align*}
		\Psi(x+y) = \|x+y + c \ones\|_p \le \|x + c\ones\|_p + \|y\|_p \le \Psi(x) + \Psi(y).
	\end{align*}
	
	For the other desired properties, we first see that $\|x\|_p \leq \Psi(x)$. This is because by convexity $\| x + c \ones \|_p - \| x \|_p \geq \langle \nabla \| x \|_p, c \ones \rangle = c \left\| \nabla \| x \|_p \right\|_1 \geq c$. Furthermore, by triangle inequality $\Psi(x) \leq \|x\|_p + \| c \ones \|_p - c$ and observe that $\| c \ones \|_p - c = \frac{(p-1)(\|\ones\|_p - 1)}{\epsilon \|\ones\|_p} \leq \frac{1}{\epsilon} \min\{p-1, \ln d\}$. So it has approximation error $(1,\min\{p-1, \ln d\})$. Next, we compute the gradients: for all $x,y \in \R^d_+$
\[
		\nabla_i \Psi(x) = \frac{(x_i+c)^{p-1}}{\| x + c \ones\|_p^{p-1}}
\quad \text{ and } \quad 
		\nabla_i \Psi(x+y) = \frac{(x_i+y_i+c)^{p-1}}{\| x + y + c \ones\|_p^{p-1}}.
\]

Observe that by triangle inequality,
\begin{align*}
\|x+y+c\ones\|_p ~\leq~ \|x+c\ones\|_p + \|y\|_p ~&=~ \|x+c\ones\|_p \left( 1 + \frac{\|y\|_p}{\|x+c\ones\|_p } \right)\\ &\leq~ \|x+c\ones\|_p \left( 1 + \frac{\|y\|_p}{c\|\ones\|_p } \right) ~\leq~ \exp\left( \frac{\|y\|_p}{c\|\ones\|_p } \right) \|x+c\ones\|_p.
\end{align*}
Therefore,
\begin{align*}
		\nabla_i \Psi(x+y) ~=~ \frac{(x_i+y_i+c)^{p-1}}{\| x + y + c \ones\|^{p-1}} ~&\geq~ \exp\left( -\frac{(p-1)\|y\|_p}{c\|\ones\|_p } \right) \frac{(x_i+c)^{p-1}}{\| x + c \ones\|^{p-1}} \\
		~&=~ \exp\left( -\e \|y\|_p \right) \nabla_i \Psi(x), 
\end{align*}
which proves its gradient-stability.
\end{proof}

	\subsection{Smoothing of Top-\texorpdfstring{$k$}{k} Norm in Polynomial Time}
	\label{sec:topk-polytime}
	
    \thmtopkpolytime*

	Recall that the existential construction for a $\delta'$-gradient stable approximation in \Cref{sec:topk} uses 
	\begin{align}
	f_\e(x) := \E_{K,\nu} \|x + \nu\|_{\mtop{K}}, ~\text{ where}
	\end{align}
	\begin{enumerate}
		\item $\nu_i$ is an independent Exponential random variable with rate $\lambda = \epsilon k$ for every $i \in [d]$.
		\item $K$ follows the geometric distribution starting at 1 with parameter $p = 1-q$ for $q = 1-\frac{\delta'}{k}$, i.e., $\Pr(K = i) = q^{i-1} (1-q)$ for $i \ge 1$ and $\E[K] = \frac{k}{\delta'}$.
		\item For $i > d$, we define $\|x\|_{\mtop{i}} = \|x\|_{\mtop{d}}$.  
	\end{enumerate}
	
	To obtain a $\delta$-gradient stable approximation in polynomial time, we approximate $\nu$ by samples and set $\delta' = \frac{\delta}{2}$. That is, for an $s$, which is appropriately chosen as a polynomial in $\frac{d}{\rho}$ to be defined later, let $V^1,\ldots,V^s \in \R^d$ be independent RV's where each coordinate $V^\ell_i$ is an independent Exponential random variable with parameter $\lambda = \e k$. Then instead of $f$ we use 
	\begin{align}
		\tilde{f}_\e(x) ~=~ \E_K \frac{1}{s} \sum_{\ell=1}^s \|x + V^\ell\|_{\mtop{K}} ~+~ \frac{2 \Delta}{\delta''} \cdot \|x\|_1, \label{eq:topkApproxPoly}
	\end{align}
	where $\delta'' = \frac{1}{2} \delta$.  Since the random variable $K$ only has $d$ scenarios, we can compute the expectation by summation. Then it is easy to see that we can indeed perform value evaluations on $\tilde{f}_\e$ in polynomial time in $s$ and $d$.

    Regarding its gradients, $\tilde{f}$ is differentiable almost-everywhere, namely for all $x$ where $x_i + V^\ell_i \neq ((x+V^\ell)_{-i})_{(j)}$ for all $i,j \in [d]$ (recall that given a vector $x$,  $x_{(i)}$ denotes its $i$th largest coordinate and $x_{-i} \in \R^{d-1}$ is the vector $x$ without its $i$th coordinate). Wherever it is differentiable, its gradient is given by 
    $$\nabla_i \tilde{f}(x) \,=\, {\textstyle \sum_{j=1}^\infty} \Pr(K=j)\cdot \frac{1}{s} \sum_{\ell=1}^s \ones\Big(x_i + V^\ell_i > ((x+V^\ell)_{-i})_{(j)}\Big) + \frac{2 \Delta}{\delta''},$$ and so can also be evaluated in polynomial time in $s$ and $d$. 
		
	\paragraph{Concentration.}
		We will argue that $\tilde{f}_\e$ fulfills the desirable properties if the draws $V^1, \ldots, V^s$ sufficiently concentrate. To this end, we will use the Dvoretzky–Kiefer–Wolfowitz inequality (DKW inequality) and Chebyshev's inequality. Both apply to distributions over the reals. We let $X, X_1,\ldots,X_s$ be independent samples all drawn from the same distribution.
		
		The DKW inequality~\cite{DKW} states that the CDF and the empirical distribution are close, namely that for any $p > 0$ with probability at least $1-p$ we have
		\begin{align*}
			\bigg|\frac{1}{s} \sum_{\ell = 1}^s \ones(X_\ell > t) ~-~ \Pr(X > t) \bigg| ~\le~ \sqrt{\frac{\ln (2/p)}{2s}}, ~~~~~ \forall t.
		\end{align*}
	
	Analogously, Chebyshev's inequality gives us a comparison of the expectation and the empirical average. It states that with probability at least $1-p$
	\begin{align*}
		\bigg|\frac{1}{s} \sum_{\ell = 1}^s X_\ell - \E[X] \bigg| ~\le~ \sqrt{\frac{\mathbf{Var}[X]}{p}}\,.
	\end{align*}

	Applying the DKW inequality on each of the components of $V^1, \ldots, V^s$ and Chebyshev's inequality on $\max_i V^1_i, \ldots, \max_i V^s_i$, by a union bound, we can assume that with probability $1 - \rho$, we have
	\begin{align}
	\label{eq:fromdkw}
	\bigg|\frac{1}{s} \sum_{\ell = 1}^s \ones(V^\ell_r > t) ~-~ \Pr(\nu_r > t) \bigg| ~\le~ \Delta \qquad \text{ for all $r \in [d]$,}
	\end{align}
	and
	\begin{align}
	\label{eq:fromchebyshev}
	\bigg|\frac{1}{s} \sum_{\ell = 1}^s \max_r V^\ell_r - \E[\max_r \nu_r] \bigg| ~\le~ \Delta',
	\end{align}
	where $\Delta = \sqrt{\frac{\ln (4 d/\rho)}{2s}}$ and $\Delta' = \sqrt{\frac{2 \mathbf{Var}[\max_r \nu_r]}{s \rho}} \leq \frac{4 d}{\epsilon k \sqrt{s \rho}}$. Note that for $s \geq \max\{ 4d^2 \ln(4d / \rho), 16 d^2 / \rho\}$, we have $2 \Delta d \leq 1$ and $k \Delta' \leq \frac{1}{\epsilon}$.
	
	\paragraph{Guarantees of the function $\tilde{f}_\e$.}	In the rest of the section, we denote the function $\tilde{f}_\e(\cdot)$ by $\tilde{f}(\cdot)$ for ease of notation.
		
	\begin{lemma} \label{lem:approxTopkPoly}
	Whenever \eqref{eq:fromchebyshev} is fulfilled, the function $\tilde{f}$ defined in \Cref{eq:topkApproxPoly} satisfies
	$$\tilde{f}(x) \ge \exp(-\delta') \cdot \|x\|_{\mtop{k}} \quad \text{and} \quad \tilde{f}(x) \le \left( 1 + \frac{1}{\delta'} + \frac{2 \Delta d}{\delta''} \right) \cdot \|x\|_{\mtop{k}} + \frac{H_d}{\delta' \epsilon} +  \frac{k \Delta'}{\delta'}.$$
	Moreover, $f$ is monotone, subadditive, and convex. 	\end{lemma}
	
	\begin{proof}
	The proof mainly follows the steps in the proof \Cref{lem:approxTopk}.
	
	The lower bound works exactly the same way. By the choice of the distributions for $K$ and monotonicity, we have
	\begin{align*}
	\tilde{f}(x) ~& =~ \E_K \frac{1}{s} \sum_{\ell=1}^s \|x + V^\ell\|_{\mtop{K}} + \frac{2 \Delta}{\delta''} \cdot \|x\|_1 ~\geq~ \E_{K} \|x\|_{\mtop{K}} ~=~ \sum_{j=1}^\infty \Pr(K=j) \|x\|_{\mtop{j}} \\
	~& \geq~ \Pr(K \geq k) \|x\|_{\mtop{k}} ~=~ q^{k-1} \|x\|_{\mtop{k}} ~=~ \left(1 - \frac{\delta'}{k}\right)^{k-1} \|x\|_{\mtop{k}} ~\geq~ \exp(-\delta') \|x\|_{\mtop{k}}.
	\end{align*}
	
	To obtain the upper bound, we again use the triangle inequality
	\[
	\tilde{f}(x) ~=~  \E_K \frac{1}{s} \sum_{\ell=1}^s \|x + V^\ell\|_{\mtop{K}} + \frac{2 \Delta}{\delta''} \cdot \|x\|_1 ~\leq~ \E_K \|x\|_{\mtop{K}} + \E_K \frac{1}{s} \sum_{\ell=1}^s \|V^\ell\|_{\mtop{K}} + \frac{2 \Delta}{\delta''} \cdot \|x\|_1.
	\]
	To bound the first term, we use again $\E_K \|x\|_{\mtop{K}} \leq \|x\|_{\mtop{k}} \left( 1 + \frac{1}{\delta'} \right)$. For the third term, we naturally have $\|x\|_1 \leq d \|x\|_{\mtop{k}}$.
	
	Finally, by \eqref{eq:fromchebyshev}, we can bound the second term using
	\[
	\E_K \frac{1}{s} \sum_{\ell=1}^s \|V^\ell\|_{\mtop{K}} \leq \E_K K \frac{1}{s} \sum_{\ell=1}^s \max_i V^\ell_i \leq \frac{k}{\delta'} \cdot \left( \E_\nu \max_i \nu_i + \Delta' \right) = \frac{k}{\delta'} \cdot \left(\frac{H_d}{\lambda} + \Delta' \right)= \frac{H_d}{\delta' \epsilon} + \frac{k \Delta'}{\delta'}.
	\]
	
	The monotonicity, subadditivity, and convexity of $f$ follows from the fact that the $\mtop{k}$ norms satisfy these properties and that $f$ is formed by taking convex combinations of these norms.
	\end{proof}
	
  Now for the gradient-stability of $\tilde{f}$.
	
	\begin{lemma} \label{thm:stabPoly}
		Whenever \eqref{eq:fromdkw} is fulfilled, for all $i$ and all non-negative $x,y \in \R_+^d$ where $\tilde{f}$ is differentiable, we have 
		\[ \nabla_i \tilde{f}(x+y) ~\ge~ \exp(-\epsilon \|y\|_{\mtop{k}}-\delta' - \delta'') \cdot \nabla_i \tilde{f}(x).
		\]
	\end{lemma}

	\begin{proof}
	We mainly follow the same steps as in the proof of \Cref{thm:stab}, including the use of \Cref{lemma:gamma}. Now, we get
	\begin{align*}
		\nabla_i \tilde{f}(x+y) \,&=\, {\textstyle \sum_{j=1}^\infty} \Pr(K=j)\cdot \frac{1}{s} \sum_{\ell=1}^s \ones\Big(x_i + y_i + V^\ell_i > ((x+y+V^\ell)_{-i})_{(j)}\Big) + \frac{2 \Delta}{\delta''} \notag \\
	 \,&\ge\, {\textstyle \sum_{j=k}^\infty} \Pr(K=j)\cdot \frac{1}{s} \sum_{\ell=1}^s \ones\Big(x_i + y_i + V^\ell_i > ((x+y+V^\ell)_{-i})_{(j)}\Big) + \frac{2 \Delta}{\delta''} \notag \\
	 \,&=\, {\textstyle \sum_{j=1}^\infty} \Pr(K=j+k-1)\cdot \frac{1}{s} \sum_{\ell=1}^s \ones\Big(x_i + y_i + V^\ell_i > ((x+y+V^\ell)_{-i})_{(j+k-1)}\Big) + \frac{2 \Delta}{\delta''} \notag\\
	 \,&=\, q^{k-1} \cdot{\textstyle \sum_{j=1}^\infty} \Pr(K=j)\cdot \frac{1}{s} \sum_{\ell=1}^s \ones\Big(x_i + y_i + V^\ell_i > ((x+y+V^\ell)_{-i})_{(j+k-1)}\Big) + \frac{2 \Delta}{\delta''} \notag	 \\
	 \,&\ge\, q^{k-1} \cdot{\textstyle \sum_{j=1}^\infty} \Pr(K=j)\cdot \frac{1}{s} \sum_{\ell=1}^s \ones\Big(x_i + y_i + V^\ell_i > ((x+V^\ell)_{-i})_{(j)} + \frac{\|y\|_{\mtop{k}}}{k} \Big) + \frac{2 \Delta}{\delta''}	\notag \\
	 \,&\ge\, q^{k-1} \cdot{\textstyle \sum_{j=1}^\infty} \Pr(K=j)\cdot \frac{1}{s} \sum_{\ell=1}^s \ones\Big(V^\ell_i > ((x+V^\ell)_{-i})_{(j)} + \frac{\|y\|_{\mtop{k}}}{k} - x_i \Big) + { \frac{2 \Delta}{\delta''}} 
	\end{align*}
	where the second inequality uses \Cref{lemma:gamma}.
	
	Now, we use the fact that \eqref{eq:fromdkw} holds. This lets us relate the empirical probabilities to the CDF of an exponential distribution. (Recall $\nu_i$ is  Exponentially distributed with parameter $\lambda$.) So, we get
	\begin{align*}
	&\frac{1}{s} \sum_{\ell=1}^s \ones\Big(V^\ell_i > ((x+V^\ell)_{-i})_{(j)} + \frac{\|y\|_{\mtop{k}}}{k} - x_i \Big) + \frac{2 \Delta}{\delta''} \\
	&~~\stackrel{\eqref{eq:fromdkw}}{\ge} \Pr\Big(\nu_i > ((x+V^\ell)_{-i})_{(j)} +\frac{\|y\|_{\mtop{k}}}{k} - x_i \Big) + \frac{2 \Delta}{\delta''} - \Delta \\
	&~~\ge \exp\bigg(- \frac{\lambda \|y\|_{\mtop{k}}}{k}\bigg) \cdot \Pr\Big(\nu_i > ((x+V^\ell)_{-i})_{(j)} - x_i \Big) + \frac{2 \Delta}{\delta''} - \Delta\\
	&~~\stackrel{\eqref{eq:fromdkw}}{\ge} \exp\bigg(- \frac{\lambda \|y\|_{\mtop{k}}}{k}\bigg) \cdot \frac{1}{s} \sum_{\ell=1}^s \ones\Big(V^\ell_i > ((x+V^\ell)_{-i})_{(j)} - x_i \Big) + \frac{2 \Delta}{\delta''} - 2 \Delta,
	\end{align*}
	where the second inequality uses the property of the exponential distribution. This then gives (using $\lambda = \e k$)
	\begin{align*}
	\nabla_i \tilde{f}(x+y) \,&\ge\, \exp(- \e \|y\|_{\mtop{k}}) \cdot q^{k-1} \cdot{\textstyle \sum_{j=1}^\infty} \Pr(K=j) \frac{1}{s} \sum_{\ell=1}^s \ones\Big(V^\ell_i > ((x+V^\ell)_{-i})_{(j)} - x_i \Big) + \frac{2 \Delta}{\delta''} - 2 \Delta\\
	\,&=\, \exp(- \e \|y\|_{\mtop{k}}) \cdot q^{k-1} \bigg(\nabla_i \tilde{f}(x) - \frac{2 \Delta}{\delta''}\bigg) + \frac{2 \Delta}{\delta''} - 2 \Delta\\
	\,&\ge\, \exp(- \e \|y\|_{\mtop{k}}) \cdot q^{k-1} \bigg(\nabla_i \tilde{f}(x) - 2 \Delta\bigg)	\\
		\,&\ge\, (1-\delta'') \exp(- \e \|y\|_{\mtop{k}}) \cdot q^{k-1} \nabla_i \tilde{f}(x),
	\end{align*}
	where the last inequality uses the fact that $\delta'' \nabla_i \tilde{f}(x) \ge 2 \Delta$.
But by our choice of $q$, we have $q^{k-1} = (1 - \frac{\delta'}{k})^{k-1} \geq \exp(-\delta')$, which finally concludes the proof of \Cref{thm:stabPoly}.
\end{proof}

	\paragraph{Proof of Theorem \ref{thm:topk-polytime}.}
We can now construct a gradient-stable approximation $\Psi_\e$ of $\mtop{k}$: set $s = \max\{ 4d^2 \ln(4d / \rho), 16 d^2 / \rho\}$ and define $\Psi_\e(\cdot) = \exp(\delta') \cdot f_\e(\cdot)$. Gradient stability of $\Psi_\e$ follows directly, and so does the fact that we can perform value and gradient evaluations in time polynomial in $\frac{d}{\rho}$. Also directly we have $\Psi_\e(\cdot) \ge \|\cdot\|_{\mtop{k}}$ and $$\Psi_\e(x) \le \exp(\delta')\left( 1 + \frac{1}{\delta'}   + \frac{2 \Delta d}{\delta''} \right) \cdot \|x\|_{\mtop{k}} + \exp(\delta' + \delta'') \frac{H_d + \Delta'}{\delta' \epsilon}.$$

If we have , then $2 \Delta d \leq 1$ and $k \Delta' \leq \frac{1}{\epsilon}$. So, $\Psi_\e$ is upper bounded by
$$\Psi_\e(x) \le \exp(\delta)\left( 1 + \frac{4}{\delta}   \right) \cdot \|x\|_{\mtop{k}} + 2 \exp(\delta) \frac{H_d + 1}{\delta \epsilon}.$$ This concludes the proof of Theorem \ref{thm:topk-polytime}.

\subsection{Composition of Norms} \label{app:normComp}

\compNorms*

		In order to prove this theorem we first need the following important upper bound on the gradient of gradient-stable approximations. It is well-know that we have $\ip{\nabla \|u\|}{v} \le \|v\|$ (wherever the norm is differentiable). The next lemma states that still approximately holds for approximations of the norm. 
		
\begin{lemma}[Gradient bound] \label{lemma:gradientBound}
	Consider a convex and subadditive function $f : \R^d_+ \rightarrow \R$ that approximates a norm $\|\cdot\|$ in the sense $$\|x\| \le f(x) \le \alpha \|x\| + \beta~~~~~~\forall x \in \R^d_+.$$ Then for all $u,v \in \R^d_+$ we have
	\[ \ip{\nabla f(u)}{v} \le \alpha \|v\|.\]
\end{lemma}

\begin{proof}
	Using subadditivity and then convexity, we have that for every $\eta \ge 0$ 
	\[ f(u) + f(\eta v) ~\ge~ f(u+ \eta v) ~\ge~ f(u) + \ip{\nabla f(u)}{\eta v},
	\]
	which gives $\ip{\nabla f(u)}{v} \le \frac{1}{\eta} f(\eta v)$. Further using the norm approximation properties of $f$ we get 
	\[ \ip{\nabla f(u)}{v} ~\le~ \frac{1}{\eta} \bigg( \alpha\|\eta v\| + \beta \bigg) ~=~ \alpha \|v\| + \frac{\beta}{\eta}.
	\]
	Taking $\eta \rightarrow \infty$ then gives the desired bound. 
\end{proof}

  We are now ready to prove \Cref{lem:composition}. 

	\begin{proof} [Proof of \Cref{lem:composition}]
		To simplify the notation, let $\alpha^* = \max_{i \ge 1} \alpha_i$. Fix $\bar{\e} > 0$, and define $\e_0 = \frac{\bar{\e}}{\alpha^* + c}$ and also $\e_i = \frac{c \bar{\e}}{\alpha^* + c}$ for $i \ge 1$. 
		
Given the assumptions, for $i \ge 0$ let $f_i$ satisfy the requirements of a $\delta_i$-gradient-stable approximation of $\|\cdot\|_{(i)}$ (\Cref{def:gradStable}) of error $(\alpha_i,\gamma_i)$ for $\e = \e_i$. Then define the function $h$ by replacing each term of $\norm(\cdot)$ by its gradient-stable approximation:
		\begin{align}
			h(x) := f_0\bigg(f_1(A_1 x),~\ldots~,f_\ell(A_\ell x)   \bigg).
			\label{eq:compose-def}
		\end{align}

	We prove that $h$ satisfies all the desired properties.

	\paragraph{$h$ is subadditive, convex, and monotone.}
	Since each $f_i$ is monotone, it follows directly that $h$ is monotone as well. For subadditivity: To simplify the notation let $F(x) = (f_1(A_1 x),\,\ldots\,,f_\ell(A_\ell x))$. By this definition $h(x) = f_0(F(x))$. Since each $f_i$ is subadditive, we have $f_i(A_i (u+v)) \le f_i(A_i u) + f_i(A_i v)$, and hence $F(u+v) \le F(u) + F(v)$ coordinate-wise. Further using the subadditivity and monotonicity of $f_0$,
	\[h(u+v) ~=~ f_0(F(u+v)) ~\le~ f_0(F(u) + F(v)) ~\le~ f_0(F(u)) + f_0(F(v)) ~=~ h(u) + h(v),\] 
	which proves subadditivity of $h$.
	
	Using the same argument, but replacing the subadditivity assumption by the convexity assumption, we see that $h$ is convex as well. 

	\paragraph{$h$ has error $(\alpha,\gamma)$.}
	Since each $f_i$ lower bounds the respective norm $\|\cdot\|_{(i)}$, it follows directly that their composition $h$ lower bounds $\norm(\cdot)$ by monotonicity of $\|\cdot\|_{(0)}$.
	
	For the upper bound, the approximation property of $f_i$ gives that $f_i(A_i u) \le \alpha_i \|A_i u\|_{(i)} + \frac{\gamma_i}{\e_i}$, and hence by the monotonicity and approximation property of $f_0$ we have
	\begin{align*}
		h(u) &\le f_0\bigg(\alpha_1 \|A_1 v\|_{(1)} + \frac{\gamma_1}{\e_1},~\ldots,~\alpha_{\ell} \|A_{\ell} v\|_{(\ell)} + \frac{\gamma_\ell}{\e_\ell} \bigg)\\
		&\le \alpha_0\cdot  \bigg\|\Big(\alpha_1 \|A_1 v\|_{(1)} + \frac{\gamma_1}{\e_1},~\ldots~,\alpha_\ell \|A_\ell v\|_{(\ell)} + \frac{\gamma_\ell}{\e_\ell}\Big)\bigg\|_{(0)} + \frac{\gamma_0}{\e_0}\\
		&\le \alpha_0\cdot  \bigg\|\Big(\alpha_1 \|A_1 v\|_{(1)},~\ldots~,\alpha_\ell \|A_\ell v\|_{(\ell)}\Big)\bigg\|_{(0)} + \alpha_0 \cdot \bigg\|\Big(\frac{\gamma_1}{\e_1},\ldots,\frac{\gamma_\ell}{\e_\ell}\Big)\bigg\|_{(0)} +  \frac{\gamma_0}{\e_0}\\
		&\le \alpha_0 \cdot \alpha^* \bigg\|\Big(\|A_1 v\|_{(1)},~\ldots~,\|A_\ell v\|_{(\ell)}\Big)\bigg\|_{(0)} + ~\frac{\alpha_0 (\alpha^* + c)}{c \bar{\e}} \cdot \|(\gamma_1,\ldots,\gamma_\ell)\|_{(0)} + \frac{\gamma_0 (a^* + c)}{\bar{\e}}\\
		&= \alpha \cdot \norm(v) + \frac{\gamma}{\bar{\e}},
	\end{align*}
	where the next-to-last inequality uses the monotonicity of $\|\cdot\|_{(0)}$, and the last inequality the definition of $\alpha$ and $\gamma$ from the statement of the theorem.
			
	\paragraph{Gradient-stability of $h$.} Let $A^j_i$ denote the $j$th column of the matrix $A_i$.	By chain rule we have
		\begin{align}
			\nabla_j h (x) = \sum_{i \ge 1} (\nabla_i f_0) (F(x)) \cdot \ip{(\nabla f_i)(A_i x)}{A^j_i},\label{eq:compose-gradient}
		\end{align}
		where, for example, $(\nabla_i f_0) (F(x))$ denotes the $i$-th partial derivative of $f_0$ evaluated at the point $F(x)$. 
		Our goal is to lower bound the right-hand side when $x = u+v$ for $u,v \in \R^n_+$.
		
		First we observe that for all $f_i$ and $u,v \ge 0$ we have $f_i(u+v) - f_i(u) \le \alpha_i \|v\|_{(i)}$: using convexity of $f_i$ and then the gradient bound from Lemma \ref{lemma:gradientBound}, we get
		\begin{align*}
			f_i(x+y) - f_i(x) ~\le~ \ip{\nabla f_i(x+y)}{y} ~\le~ \alpha_i \|y\|_{(i)}.
		\end{align*}
		Employing this on all the coordinates of $F$ and using the monotonicity of $\|\cdot\|_{(0)}$, we then get
		\begin{align*}
			\|F(u+v) - F(u)\|_{(0)} ~&\le~ \bigg\|\Big(\alpha_1 \|A_1 v\|_{(1)},~\ldots~, \alpha_\ell \|A_\ell v\|_{(\ell)}\Big) \bigg\|_{(0)} \\
			~&\le~ \bigg\|\Big(\alpha^* \|A_1 v\|_{(1)},~\ldots~, \alpha^* \|A_\ell v\|_{(\ell)}\Big) \bigg\|_{(0)} ~=~ \alpha^* \cdot \norm(v).
		\end{align*}
		So using the gradient-stability  of $f_0$ we have for all $i$
		\begin{align*}
			(\nabla_i f_0)(F(u+v)) &\ge \exp\bigg(-\e_0 \|F(u+v) - F(u)\|_{(0)} - \delta_0 \bigg) \cdot \nabla f_0(F(u)) \\
			&\ge \exp\bigg(-\e_0 \cdot \alpha^* \cdot \norm(v)  - \delta_0 \bigg) \cdot (\nabla_i f_0)(F(u)).
		\end{align*}
		Furthermore, the gradient-stability of each $f_i$ implies 
		\[
		(\nabla f_i)(A_i (u+v)) ~\geq~ \exp(-\e_i \|A_i v\|_{(i)} - \delta_i) (\nabla f_i)(A_i u).
		\]
		Because all entries in the matrices $A_i$ as well as the gradients $\nabla f_i$ are non-negative, we can apply these two bounds on \eqref{eq:compose-gradient} and get
		\begin{align}
		&\nabla_j h (u+v) \notag\\ 
		&\ge \sum_{i} \exp\bigg(-\e_0 \cdot \alpha^* \cdot \norm(v)  - \delta_0 \bigg) \cdot (\nabla_i f_0)(F(u)) \cdot \exp(-\e_i \|A_i v\|_{(i)} - \delta_i) \cdot \ip{(\nabla f_i)(A_i u)}{A^j_i}  \notag\\
		&\ge \exp\bigg(-\e_0 \cdot \alpha^* \cdot \norm(v)  - \delta_0 - \max_{i \ge 1} (\e_i \|A_i v\|_{(i)} + \delta_i) \bigg) \cdot \sum_i (\nabla_i f_0)(F(u)) \cdot \ip{(\nabla f_i)(A_i u)}{A^j_i}  \notag\\
		&= \exp\bigg(-\e_0 \cdot \alpha^* \cdot \norm(v)  - \delta_0 - \max_{i \ge 1} (\e_i \|A_i v\|_{(i)} + \delta_i) \bigg) \cdot \nabla_j h(u).   \label{eq:comp}
		\end{align}
	Moreover, given the assumption that $\|\cdot\|_{\infty} \le \|\cdot\|_{(0)}$, we have 
	\begin{align*}
	\max_{i \ge 1} (\e_i \|A_i v\|_{(i)} + \delta_i) &\le \max_{i \ge 1} \e_i \|A_i v\|_{(i)} + \max_{i \ge 1} \delta_i \\
	&\le \Big\|\big(\e_1 \|A_1 v\|_{(1)},\ldots,\e_\ell \|A_\ell v\|_{(\ell)}\big)\Big\|_{(0)} + \max_{i \ge 1} \delta_i\\
	&\le \frac{c \bar{\e}}{\alpha^* + c} \cdot \norm(v) + \max_{i \ge 1} \delta_i.
	\end{align*}
	Plugging this in \eqref{eq:comp} and using the definitions $\e_0 = \frac{\bar{\e}}{\alpha^* + c}$ and $\delta = \delta_0 + \max_{i \ge 1} \delta_i$ gives
		\begin{align*}
		\nabla_j h (u+v) \ge \exp\Big(- \bar{\e} \cdot \norm(v)  - \delta \Big) \cdot \nabla_j h(u),
		\end{align*}	
		proving the gradient-stability of $h$.
	
	\focs{\paragraph{Computation.} Note that the function $h$ is given explicitly by \eqref{eq:compose-def} and its gradient $\nabla h$ is given explicitly by \eqref{eq:compose-gradient}. Therefore, both any function value and any gradient can be computed efficiently given oracle access to the norm approximations being used in the composition.}
	
	Since these hold for every $\bar{\e} > 0$, we see that $\norm(\cdot)$ admits a $\delta$-gradient-stable approximation with error $(\alpha,\gamma)$. This concludes the proof of \Cref{lem:composition}.
	\end{proof}

\subsection{Explicit Representation of a Symmetric Norm in  Polynomial Time} \label{sec:SymmetricNormExplicit}

In this section we will use \ballopt to get an explicit approximate representation of a symmetric norm.

\SymmetricNormExplicit*

\begin{proof}
For any symmetric monotone norm $\|\cdot \|$, define set $\calX := \{ x \in \R_+^d \mid \max_{v:\|v\|\leq 1} \langle x, v \rangle \leq 1 \}$. We know  that $\|v\| = \max_{x \in \calX} \langle x, v \rangle$ for any vector $v \in \R^d$. Also,  monotonicity implies that all vectors in $\calX$ are non-negative and symmetry implies that for any vector $x \in \calX$ all  $d!$ permutations of $x$ are also inside $\calX$. 


We first simplify the set $\calX$. Define 
$\overline{\calX}:= \big\{\overline{x} \mid \exists x \in \calX \text{ s.t. } \overline{x}_i = x_i \cdot \one\big[x_i \in [\eps'\frac{\|e_1\|}{d}, \|e_1\|] \big] \text{ for all } i\in \{1,\ldots, d\} \big\}.$ That is, $\overline{\calX}$ consists of vectors in $\calX$ where we $0$ any coordinates outside  $[\eps'\frac{\|e_1\|}{d}, \|e_1\|]$.

\begin{claim} \label{claim:sparserSet}
For any $v\in \R^d$, we have
\[ \max_{x \in \overline{\calX}} \langle x, v \rangle ~\leq~ \|v\| ~\leq~ (1+\eps') \max_{x \in \overline{\calX}} \langle x, v \rangle.
\]
\end{claim}
\begin{proof}
The first inequality follows since $\max_{x \in {\calX}} \langle x, v \rangle = \|v\|$ and $\overline{\calX}$ is only formed by zeroing some of the coordinates of non-negative vectors of $\calX$.

For the second inequality, we can first assume that $v_1 \geq v_2 \geq \ldots \geq v_d$ since the norm $\|\cdot\|$ is symmetric. Now 
consider $x_* \in \calX$ that achieves $\|v\|= \langle x_* , v \rangle$, where symmetry again implies that $x_*$ has non-increasing coordinates: $x_*(1) \geq \ldots \geq x_*(d)$. We will show that the vector $\overline{x}_* \in \overline{\calX}$ corresponding to $x_*$ satisfies $\|v\| \leq (1+\eps') \langle\overline{x}_* , v \rangle$.
First, observe that $x_*(1) \leq  \|e_1\|$
since 
\[ 1 ~\geq~ \max_{v: \|v\|\leq 1} \langle x_*, v \rangle ~\geq~ \Big\langle x_*, \frac{e_1}{\|e_1\|} \Big\rangle ~=~ \frac{x_*(1)}{\|e_1\|} .
\]
Thus, all coordinates of $x_*$ are at most $\|e_1\|$ and are not zeroed in $\overline{x}_*$. Since coordinate of $x_*$ that is zeroed in   $\overline{x}_*$ is at most $\eps'\frac{\|e_1\|}{d}$, we have
\[ \langle v , \overline{x}_* \rangle ~\geq ~ \langle v , x_* \rangle - \Big\langle v , \sum_i e_i \cdot \eps'\frac{\|e_1\|}{d} \Big\rangle ~\geq~ \|v\| - \eps' \cdot v_1 \|e_1\|  ~\geq~ \|v\| - \eps' \cdot \|v\|,
\]
where the last inequality uses $\|v\| \geq \|v_1 e_1 \|$.
\end{proof}

Next we sparsify the vectors in $\overline{\calX}$ as in \cite{CS-STOC19} to obtain set $\calX'$. Consider any vector  $\overline{x} \in \overline{\calX}$ with $\overline{x}_1 \geq \overline{x}_2 \geq \ldots \geq \overline{x}_d$.  We first define vector $\widetilde{x}$ corresponding to $\overline{x}$ where coordinate change $\widetilde{x}_i>\widetilde{x}_{i+1}$ is only possible when $i$ is a power of $2$. (We will only prove the weaker result where we lose an extra factor of $2$. But to make this factor smaller, we should consider  powers of $1+\delta$ (with floors)). Formally, define $\widetilde{x}_i = x_i$ if $i = \min\{2^s,d\}$ for some integer $s \geq 0$, and otherwise if $s$ is the unique integer with $2^{s-1} < i < 2^{s}$ then $\widetilde{w}_i = w_{\min\{2^s,d\}}$. 

\begin{claim}[Claim 4.1 in \cite{CS-STOC19}]
For any vector $v \in \R_+^d$ with non-increasing coordinates we have $\langle v, \widetilde{x} \rangle \leq \langle v, x \rangle \leq 2 \langle v, \widetilde{x}  \rangle$.
\end{claim}

Next, we obtain $x'$ from $\widetilde{x}$ by  rounding down each coordinate of  $\widetilde{x}$  
to the nearest power of $(1+\eps')$. The set $\calX'$ is now defined by taking the union of all such vectors $x'$ and their $d!$ permutations.
%
Combining with \Cref{claim:sparserSet}, this implies
\begin{align} \label{eq:propOfXPrime}
\max_{x \in \calX'} \langle x, v \rangle ~\leq~ \|v\| ~\leq~ 2 (1+\eps')^2 \max_{x \in \calX'} \langle x, v \rangle.
\end{align}

Observe that the number of distinct $x' \in \calX'$ with $x'_1 \geq x'_2 \geq \ldots \geq x'_d$  is polynomially bounded. To prove this, we need the following claim.

\begin{claim}[Claim 5.3 in \cite{CS-STOC19}] \label{claim:CS5.3}
There are at most $(2e)^{\max\{N,k\}}$  non-increasing sequences of $k$ integers chosen from $\{ 0, \ldots, N\}$.
\end{claim}

Recall that each coordinate of $x'$ is a power of $(1+\eps')$ and is in the range $[\frac{\eps'}{1+\eps'}\frac{\|e_1\|}{d}, \|e_1\|]$. Thus the number of distinct values for a coordinate of $x'$ is  $O\big( \frac{\log (d/2 \eps')}{\eps'} \big)$. Moreover, since coordinates in $x'$ only change when they are a power of $2$, we only need a non-increasing sequence of $O({\log d})$ length to define $x'$. So, by \Cref{claim:CS5.3} the number of such $x'$ is at most $(2e)^{\max\{ O( \frac{\log (d/2 \eps')}{\eps'}) ,O(\log d)\}} = \poly(d/\eps')^{1/\eps'}$.

Finally, we can define $\W'$. For  any vector $x' \in \calX'$ with $x'_1 \geq x'_2 \geq \ldots \geq x'_d$, we define a weight vector $w' \in \W'$ as follows: starting with $w'_d := x'_d$ let $w'_i := x'_i - x'_{i+1}$ for $i \in \{1,\ldots, d-1\}$. The main observation is that for any vector $v$ with $v_1 \geq v_2 \geq \ldots \geq v_d$, 
\[ \langle v , x' \rangle ~=~ \sum_{i=1}^d w'_i \|v\|_{\mtop{i}}.
\]
 Thus, combining with \eqref{eq:propOfXPrime}, we get
\begin{align} \label{eq:propOfWPrime}
\max_{w \in \W'}\sum_{i=1}^d w'_i \|v\|_{\mtop{i}} ~\leq~ \|v\| ~\leq~ 2(1+\eps')^2 \max_{w \in \W'}\sum_{i=1}^d w'_i \|v\|_{\mtop{i}}.
\end{align}

Moreover,  the size of $\W'$ is polynomially bounded by construction. It's also easy to find $\W'$ efficiently since we can test for any of these polynomially many candidate weight vectors $w'$ whether they belong to $\W'$: use the \ballopt oracle to test whether $\max_{v: \|v\|\leq 1} \langle x', v \rangle \leq 1$, where $x'$ is the unique vector corresponding to $w'$ with $x'_1 \geq x'_2 \geq \ldots \geq x'_d$ as defined in the construction.
\end{proof}


{\small
\bibliographystyle{alpha}
\bibliography{bib,online-lp-short3}
}

\end{document}